\documentclass{llncs}
\usepackage[utf8]{inputenc}
\usepackage{cite}
\usepackage{defs}
\usepackage{graphicx}
\usepackage{mathpartir}
\usepackage{multirow}
\usepackage{thm-restate}
\usepackage{url}

\newcommand{\mytitle}{View-Based \owickigries{} Reasoning for Persistent x86-TSO (Extended Version)}
\title{\mytitle\thanks{Vafeiadi Bila is supported by VeTSS. Dongol is supported by EPSRC
    grants EP/V038915/1, EP/R032556/1, EP/R025134/2 and ARC
    Discovery Grant DP190102142.
    Lahav is supported by the Israel Science
Foundation (grant 1566/18), by the European Research Council
(ERC) under the European Union’s Horizon 2020 research and innovation programme
(grant agreement no. 851811), and by the
Alon Young Faculty Fellowship. 
Raad is supported by a UKRI Future Leaders Fellowship [grant number MR/V024299/1].
Wickerson is supported by an EPSRC Programme Grant (EP/R006865/1).}}

\author{Eleni Vafeiadi Bila\inst{1}\orcidID{0000-0003-3399-0736} \and Brijesh Dongol\inst{1}\orcidID{0000-0003-0446-3507} \and Ori Lahav\inst{2}\orcidID{0000-0003-4305-6998} \and Azalea Raad\inst{3}\orcidID{0000-0002-2319-3242} \and John Wickerson\inst{3}\orcidID{0000-0001-6735-5533}}
\institute{University of Surrey, Guildford, UK \and Tel Aviv University, Tel Aviv, Israel \and Imperial College London, London, UK}

\begin{document}

\maketitle



\begin{abstract}

  The rise of persistent memory is disrupting computing to its core.
  Our work aims to help programmers navigate this brave new world by
  providing a program logic for reasoning about x86 code that uses
  low-level operations such as memory accesses and fences, as well as persistency primitives such as flushes.
  Our logic, \logicname{}, benefits from a simple underlying
  operational semantics based on \emph{views}, is able to handle
  \emph{optimised} flush operations, 
  and is mechanised in the Isabelle/HOL proof
  assistant.  We detail the proof rules of \logicname{}
  and prove them sound. We also show how \logicname{} can be used to
  reason about a range of challenging single- and multi-threaded persistent programs.

\end{abstract}

\section{Introduction}
\label{sec:intro}
In our era of big data, the long-established boundary between `memory'
and `storage' is increasingly blurred. Persistent memory is
a technology that sits in both camps, promising both the durability of disks 
and data access times similar to those of DRAM.
Embracing this technology requires rethinking our decades-old
programming paradigms. As data held in memory is no longer wiped
after a system restart, there is an opportunity to write
\emph{persistent} programs -- programs that can recover their progress
and continue computing even after a crash.

However, writing persistent programs is extremely challenging, as it
requires the programmer to keep track of which memory writes have
become persistent, and which have not. This is further complicated in
a multi-threaded setting by the intricate interplay between the rules
of memory \emph{persistency} (which determine the order in which
writes become persistent) and those of memory \emph{consistency}
(which determine what data can be observed by which threads).

To address this difficulty, we provide a foundation for persistent
programming. We develop a program logic, \logicname{}, for reasoning
about x86 code that uses low-level operations such as memory accesses
and fences, as well as persistency primitives such as flushes.  We
demonstrate the utility of \logic by using it to reason about a range
of challenging single- and multi-threaded persistent programs,
including some that demonstrate the subtle interplay between optimised
flush (\fo) and store fence (\sfence) instructions.  Using the
Isabelle/HOL proof assistant, we have mechanised the \logic rules and
proved them sound with respect to an operational semantics for x86
persistency~\cite{DBLP:conf/pldi/ChoLRK21}.  One benefit of our
Isabelle/HOL formalisation is that \logic is already partially
automated: once the user has produced a proof outline (\ie annotated
each instruction with a postcondition), they can simply use
Isabelle/HOL's \emph{sledgehammer}, which automatically decides which
axioms and rules of the proof system need invoking to verify the whole
program. Our mechanisation, which includes all the example programs
discussed in this paper, is available as auxiliary
material~\cite{Vafeiadi-Bila2022}.

\noindent\textbf{\emph{State of the art}}
To our knowledge, the only program logic for persistent programs 
is POG (Persistent \owickigries)~\cite{DBLP:journals/pacmpl/RaadLV20}. 
As with \logicname{}, POG enables reasoning about persistent x86 programs and is based on the
\owickigries{} method~\cite{DBLP:journals/acta/OwickiG76}. However,
unlike \logicname{}, POG is not mechanised in a proof assistant, and
does not support optimised flush (\fo) instructions. Optimised
flush instructions are an important persistency primitive as they are considerably faster than
ordinary flush instructions. Indeed, Intel's experiments on their
Skylake microarchitecture indicate that they can be \emph{nine times} faster
when applied to buffers that hold tens of kilobytes of
data~\cite[p.~289]{intel_optimization_manual}, and hence programmers
are impelled, ``If \fo{} is available, use \fo{} over \fl{}.''
However, \fo{} is a tricky instruction for programmers and program logic designers
alike: compared to \fl, \fo can be reordered with more
instructions under x86.

\logicname{} can reason efficiently about x86 persistency
(including \fo instructions) thanks to two
key recent advances:
\begin{enumerate*}
\item \pxes~\cite{DBLP:conf/pldi/ChoLRK21}, the view-based operational semantics of x86 persistency;
  and
\item the C11 Owicki-Gries logic~\cite{DBLP:journals/corr/abs-2004-02983,DBLP:journals/darts/DalvandiDDW20,DBLP:conf/ecoop/DalvandiDDW19} to reason about view-based operational semantics, which we adapt to \pxes.
\end{enumerate*}


\vspace{2pt}\noindent\textbf{\emph{Our contributions}}
\begin{enumerate*}

\item We present a program logic, called \logic, for reasoning about persistent x86 programs.

\item We mechanise (and partially automate) \logicname{} in Isabelle/HOL, and prove it sound relative to an established operational semantics for x86 persistency.

\item We demonstrate the utility of \logic by using it to verify several idiomatic persistent x86 programs.

\end{enumerate*}

\vspace{2pt}\noindent\textbf{\emph{Outline}}
We begin with an overview of memory consistency and persistency in x86 and provide an example-driven account of \logic reasoning (\cref{sec:motivation}). 
We describe the assertion language and proof rules of \logic in \cref{sec:proof-rules-and-og},
and verify a selection of programs using \logic in \cref{sec:example-proofs}. 
We present the view-based operational semantics of x86 persistency and 
prove the soundness of \logicname{} in \cref{sec:soundness}. 

\vspace{2pt}\noindent\textbf{\emph{Auxiliary material}}
Our Isabelle/HOL mechanisation is available as auxiliary material~\cite{Vafeiadi-Bila2022}.








\section{Overview and Motivation} 
\label{sec:motivation}

Recent operational models for weak memory use {\em views} to capture
relaxed behaviours of concurrent
programs~\cite{DBLP:conf/popl/KangHLVD17,DBLP:conf/pldi/ChoLRK21,DBLP:conf/ecoop/DalvandiDDW19,igps}, where the memory records
the entire history of writes that have taken place thus far. This way,
different threads can have different subsets of these writes (\ie
different {\em views}) visible to them.  In what follows, we review
\pxes, a view-based operational semantics for x86 persistency
(\cref{sec:pxes-at-glance}); we then 
describe \logic using a series of running examples.


\subsection{\pxes at a Glance}
\label{sec:pxes-at-glance}
In the literature of concurrency semantics, \emph{consistency} models describe the permitted behaviours of programs by constraining the volatile memory order, \ie the order in which memory writes are made visible to other threads, while \emph{persistency} models describe the permitted behaviours of programs upon recovering from a crash (\eg a power failure)
by defining the persistent memory order, \ie the order in which writes are committed to persistent memory.
To distinguish between the two, memory \emph{stores} are differentiated from memory \emph{persists}: the former denotes the process of making a write visible to other threads,
whilst the latter denotes the process of committing writes to persistent memory (durably).

\vspace{2pt}
\noindent\textbf{\emph{\pxes Consistency}}
The consistency semantics of \pxes is that of the well-known TSO (total store ordering) \cite{tso} model, where later (in program order) reads can be reordered before earlier writes on different locations. This is illustrated in the \emph{store buffering} (\ref{ex:sb}) example below (left):
\vspace{-15pt}\\
\begin{tabular}{@{} p{5.3cm} @{\hspace{30pt}} p{5.5cm} @{}}
\begin{equation}
\begin{array}{c}
\inarrII{
	\store {x}{1};\;\; \\
	\load {a} {y}
}{
	\store{y}{1};\\
	\load {b} x
} \\
a = 0 \land  b = 0: \text{\cmark}
\end{array}
\tag{\textsc{sb}}
\label{ex:sb}
\end{equation}
&
\begin{equation}
\begin{array}{c}
\inarrII{
	\store{x}{42};\;\; \\
	\store{y}{7}
}{
	\load{a}{y};\\
	\load {b} x
} \\
a = 7 \land  b = 0: \text{\xmark}
\end{array}
\tag{\textsc{mp}}
\label{ex:mp}
\end{equation}
\end{tabular}
\vspace{-10pt}\\
Specifically, assuming $x \!=\! y \!=\! 0$ initially, since $\load {a} y$ (\resp $\load {b} x$) can be reordered before $\store x 1$ (\resp $\store y 1$), it is possible to observe the weak behaviour $a \!=\! 0 \land b \!=\! 0$.
A well-known way of modelling such reorderings in TSO is through \emph{store buffers}: when a thread $\tid$ executes a write $\store{x} v$, its effects are not immediately made visible to other threads; rather they are delayed in a thread-local (store) buffer only visible to $\tid$, and propagated to the memory at a later time, whereby they become visible to other threads. For instance, when $\store{x} 1$ and $\store{y} 1$ are delayed in the respective thread buffers (and thus not visible to one another), then $\load{a} y$ and $\load{b} x$ may both read $0$.

Cho \etal~\cite{DBLP:conf/pldi/ChoLRK21} capture this by associating each thread $\tid$ with a \emph{coherence view} (also called a thread-observable view), describing the writes observable by $\tid$. 
Distinct threads may have different coherence views. For instance, after executing $\store{x} 1$ and $\store{y} 1$, the coherence view of the left thread may include $\store{x} 1$ and \emph{not} $\store{y} 1$, while that of the right may include $\store{y} 1$ and \emph{not} $\store{x} 1$. This way, $\load{a} y$ (\resp $\load{b} x$) may read the initial value $0$, as its coherence view does not include $\store{y} 1$ (\resp $\store{x} 1$).

After SC (sequential consistency) \cite{sc}, TSO is one of the strongest consistency models and supports synchronisation patterns such as \emph{message passing}, 
as shown in \ref{ex:mp} above (right), where $a \!=\! 7  \land  b \!=\! 0 $ cannot be observed. 
Specifically, (assuming $x \!=\! y \!=\! 0$ initially) if the right thread reads $7$ from $y$ (written by the left thread), then the left thread passes a message to the right. 
Under TSO, message passing ensures that the instruction writing the message and all those ordered before it (\eg $\store{x} {42}; \store{y} 7$) are executed (ordered) before the instruction reading it (\eg $\load{a} y$).
As such, since $\load{b} x$ is executed after $\load{a} y$, if $a \!=\!7$ (\ie $\store{x} {42}$ is executed before $\load{a} y$), then $b \!=\! 42$. 

\vspace{2pt}
\noindent\textbf{\emph{\pxes Persistency}}
Cho \etal \cite{DBLP:conf/pldi/ChoLRK21} recently developed the \pxes model, a view-based description of the \intelname persistency semantics, which follows a \emph{buffered}, \emph{relaxed} persistency model.
Under a buffered model, memory persists occur \emph{asynchronously}~\cite{persist-buffering}: they are buffered in a queue to be committed to persistent memory at a future time.
This way, persists occur after their corresponding stores and as prescribed by the persistency semantics, while allowing the execution to proceed ahead of persists. 
As such, after recovering from a crash, only a \emph{prefix} of the persistent memory order may have persisted.
(The alternative is \emph{unbuffered} persistency in which stores and persists happen simultaneously.)

Under relaxed persistency, the volatile and persistent memory orders may disagree: the order in which the writes are made visible to other threads may differ from the order in which they are persisted.
(The alternative is \emph{strict} persistency in which the volatile and persistent memory orders coincide.)

The relaxed and buffered persistency of \pxes is shown in \cref{subfig:litmus_tests_out_of_order_persist}.
If a crash occurs during (or after) the execution of \cref{subfig:litmus_tests_out_of_order_persist}, at crash time either write may have persisted and thus $x, y \!\in\! \{0, 1\}$ upon recovery. 
Note that the two writes cannot be reordered under \intelname (TSO) consistency and thus at no point during the normal (non-crashing) execution of \cref{subfig:litmus_tests_out_of_order_persist} is $x {=} 0, y {=} 1$ observable.
Nevertheless, in case of a crash it is possible to observe $x {=} 0, y {=} 1$ after recovery.
That is, due to the relaxed persistency of $\pxes$, the store order ($x$ before $y$) is separate from the persist order ($y$ before $x$).
More concretely, under \pxes the writes may persist
\begin{enumerate*}
	\item in any order, when they are on distinct locations; or 
	\item in the volatile memory order, when they are on the same location.%
\end{enumerate*}%
\footnote{%
Given a \emph{cache line} (a set of locations), writes on distinct cache lines may persist in any order, while writes on the same cache line persist in the volatile memory order.
For brevity, we assume that each cache line contains a single location, thus forgoing the need for cache lines. However, it is straightforward to lift this assumption.\label{footnote:cache_lines}
}

\begin{figure}[t]
\newcommand{\mymargin}{\hspace{3pt}}
\begin{tabular}{| @{\mymargin} c @{\mymargin}| @{\mymargin} c @{\mymargin} | @{\mymargin} c @{\mymargin} | @{\mymargin} c @{\mymargin} | @{\mymargin} c @{\mymargin} |}
\hline\vspace{5pt}
\begin{subfigure}[b]{0.125\textwidth}
\centering
$
\begin{array}{@{} l @{}}
	\store{x}{1};\\
	\store{y}{1}
\end{array}
$\vspace{-5pt}
\caption{}
\label{subfig:litmus_tests_out_of_order_persist}
\end{subfigure}
&
\begin{subfigure}[b]{0.125\textwidth}
\centering
$
\begin{array}{@{} l @{}}
	\store{x}{1}; \\
	\flC x; \\
	\store{y}{1}
\end{array}
$ \vspace{-5pt}
\caption{}
\label{subfig:litmus_tests_w_fl}
\end{subfigure}
&
\begin{subfigure}[b]{0.125\textwidth}
\centering
$
\begin{array}{@{} l @{}}
	\store{x}{1}; \\
	\foC x; \\
	\store{y}{1}
\end{array}
$
\vspace{-5pt}\caption{}
\label{subfig:litmus_tests_w_fo}
\end{subfigure}
&
\begin{subfigure}[b]{0.125\textwidth}
\centering
$
\begin{array}{@{} l @{}}
	\store{x}{1}; \\
	\foC x; \\
	\sfenceC; \\
	\store{y}{1}
\end{array}
$
\vspace{-5pt}\caption{}
\label{subfig:litmus_tests_w_fo_sf}
\end{subfigure}
&
\begin{subfigure}[b]{0.27\textwidth}
\vspace{5pt}
\centering
$\inarrII{
	\store{x}{1};\!\!\\
	\flC x; \!\!\\
	\store{y}{1}
}{
	\!\!\load{a}{y};\\
	\!\!\ifC {a {=} 1}\\
	\;\store{z} 1
}
$
\vspace{-5pt}\caption{}
\label{subfig:litmus_tests_fl_concurrent}
\end{subfigure}
\vspace{-8pt}\\
\hline 
\color{reccol} \rectext:\!\! $x, y \!\in\! \{0, 1\}$
& \color{reccol} \rectext:\!\! $y {=} 1 \Rightarrow x {=} 1$
& \color{reccol} \rectext:\!\! $x, y \!\in\! \{0, 1\}$
& \color{reccol} \rectext:\!\! $y {=} 1 \Rightarrow x {=} 1$
& \color{reccol} \rectext: $z {=} 1 \Rightarrow x {=} 1$ 
\\
\hline 
\end{tabular} 
%
\caption{Example \pxes programs and possible values after recovery from a crash (\rectext).
In all examples $x$, $y$, $z$ are distinct locations in persistent memory such that $x {=} y {=} z {=} 0$ initially, and $a$ is a (thread-local) register.
}
\vspace{-15pt}
\label{fig:litmus_tests}
\end{figure}

To afford more control over when pending writes are persisted, \intelname provides explicit \emph{persist} instructions such as $\flC x$ and $\foC x$ that can be used to persist the pending writes on $x$.%
\footnote{%
Executing $\flC x$ or $\foC x$ persists the pending writes on \emph{all locations in the cache line of $x$}. However, as discussed, we assume cache lines contain single locations.}
This is illustrated in \cref{subfig:litmus_tests_w_fl}: executing $\flC x$ persists the earlier write on $x$ (\ie $\store{x} 1$) to memory.
As such, if the execution of \cref{subfig:litmus_tests_w_fl} crashes and upon recovery $y {=} 1$, then $x {=} 1$. 
That is, if $\store{y} 1$ has executed and persisted before the crash, then so must the earlier $\store{x} 1; \flC x$.
Note that $y {=} 1 \Rightarrow x {=} 1$ describes a \emph{crash invariant}, in that it holds upon crash recovery \emph{regardless} of when (\ie at which program point) the crash may have occurred.
Observe that this crash invariant is guaranteed thanks to the ordering constraints on \fl instructions. Specifically, \fl instructions are ordered with respect to all writes; as such, $\flC x$ in \cref{subfig:litmus_tests_w_fl} cannot be reordered with respect to either write, and thus upon recovery $y {=} 1 \Rightarrow x {=} 1$.

However, instruction reordering means that persist instructions may not execute at the intended program point and thus not guarantee the intended persist ordering. 
Specifically, $\foC x$ is only ordered with respect to earlier writes on $x$, and may be reordered with respect to later writes, as well as earlier writes on different locations. 
This is illustrated in \cref{subfig:litmus_tests_w_fo}: $\foC x$ is not ordered with respect to $\store{y} 1$ and may be reordered after it. 
Therefore, if a crash occurs after $\store{y} 1$ has executed and persisted but before $\foC x$ has executed, then it is possible to observe $y {=} 1, x {=} 0$ on recovery.
That is, there is no guarantee that $\store{x} 1$ persists before $\store{y} 1$, \emph{despite} the intervening $\foC x$.

In order to prevent such reorderings and to strengthen the ordering constraints between $\fo$ and later instructions, one can use either \emph{fence} instructions, namely $\sfenceC$ (store fence) and $\mfenceC$ (memory fence), or atomic \emph{read-modify-write} (RMW) instructions such as compare-and-set (\kwc{CAS}) and fetch-and-add (\kwc{FAA}). 
More concretely, $\sfenceC$, $\mfenceC$ and RMW instructions are ordered with respect to all (both earlier and later) $\fo$, $\fl$ and write instructions, and can be used to prevent reorderings such as that in \cref{subfig:litmus_tests_w_fo}.
This is illustrated in \cref{subfig:litmus_tests_w_fo_sf}.
Unlike in \cref{subfig:litmus_tests_w_fo}, the intervening $\sfenceC$ ensures that $\fo$ in \cref{subfig:litmus_tests_w_fo_sf} is ordered with respect to $\store{y} 1$ and cannot be reordered after it, ensuring that $\store{x} 1$ persists before $\store{y} 1$ (\ie $y {=} 1 \Rightarrow x {=} 1$ upon recovery), as in \cref{subfig:litmus_tests_w_fl}.
Note that replacing $\sfenceC$ in \cref{subfig:litmus_tests_w_fo_sf} with $\mfenceC$ or an RMW yields the same result. 
Alternatively, one can think of $\foC x$ executing \emph{asynchronously}, in that its effect (persisting $x$) does not take place immediately upon execution, but rather at a later time. 
However, upon executing a barrier instruction (\ie \mfenceC, \sfenceC or an RMW), execution is blocked until the effect of earlier \fo instructions take place; that is, executing such barrier instructions ensures that earlier \fo behave \emph{synchronously} (like \fl).


%

%
The example in \cref{subfig:litmus_tests_fl_concurrent} illustrates how message passing can  impose persist orderings on the writes of \emph{different} threads. 
(Note that the program in the left thread of \cref{subfig:litmus_tests_fl_concurrent} is that of \cref{subfig:litmus_tests_w_fl}.)
As in \ref{ex:mp}, if $a=1$, then $\store{x} 1; \flC x$ is executed before $\load{a} y$ (thanks to message passing). Consequently, since $\store{z} 1$ is executed after $\load{a} y$ when $a = 1$, we know $\store{x} 1; \flC x$ is executed before $\store{z} 1$.
Therefore, if upon recovery $z {=} 1$ (\ie $\store{z} 1$ has persisted before the crash), 
then $x {=} 1$ ($\store{x} 1; \flC x$ must have also persisted before the crash).
As before, replacing $\flC x$ in \cref{subfig:litmus_tests_fl_concurrent} with $\foC x; C$ yields the same result upon recovery when $C$ is an $\sfenceC$/$\mfenceC$ or an RMW.

%
%
%
    

   
%
%
%
%
%
%
%


\subsection{\logic: View-Based \owickigries{} Reasoning for \pxes}

\begin{figure}[t]
\begin{center}
\begin{tabular}{c @{\hspace{5pt}} || @{\hspace{5pt}} c}
\multicolumn{2}{c}
{$\nameassert{P}{
	a = b = 0 \land\for{\tid \in \{1, 2\}} \View{x}{\tid} = \View{y}{\tid} = \{0\}
}$}\\
$
\begin{array}{@{} l @{}}
	\nameassert{P_1}{7 \notin \valspred{y}{2} \ \land\  a=0} \\
	\qquad \store{x} {42}; \annotate{\rulename{SP_1}, \rulename{Cons}} \\
	\nameassert{P_2}{\tightshade{$\View x 1 = \{42\}$} \land 7 \notin \valspred{y}{2}} \\
	\qquad \store{y}  7 ; \annotate{\rulename{SP_1}, \rulename{Cons}} \\
	\nameassert{P_3}{\true}
\end{array}
$   
& 
$
\begin{array}{@{} l @{}}
	\nameassert{Q_1}{ 
		\valspred{y}{2} \subseteq \{0, 7\} 
		\land (7 \in \valspred{y}{2} \imp \csetvalpred{y}{7}{x}{2} = \{42\} )
	} \\
	\qquad \load{a} y;  \annotate{\rulename{LP_2}}\\
	\nameassert{Q_2}{\tightshade{$
		a \in \{0, 7\} 
		\land (a = 7 \imp \valspred{x}{2} = \{42\})
	$}} \\
	\qquad \load{b}  x; \annotate{\rulename{LP_1}, \rulename{Cons}} \\	
	\nameassert{Q_3}{a = 7 \imp  \tightshade{$b = 42$}}
\end{array}
$  \\
\multicolumn{2}{c}{
\hspace{-70pt}$\nameassert{Q}{a = 7 \imp  b = 42}$
} 
\end{tabular} \vspace{-5pt}
\end{center}
\hrule
\caption{A \logic proof sketch of message passing (\ref{ex:mp}), where the \annotate{\!\!\!\!} annotation at each step identifies the \logic proof rule (in \cref{subsec:proof-rules}) applied, and the \tightshade{highlighted} assertions capture the effects of the preceding instruction. 
}
\label{fig:mp_proof}
\end{figure}


\noindent\textbf{\emph{Sequential Reasoning about Consistency using Views}}
In \cref{fig:mp_proof} we present a \logic proof sketch of \ref{ex:mp}. 
Recall that in order to account for possible write-read reorderings on \intelname architectures, \pxes associates each thread $\tid$ with a coherence view, describing the writes visible to $\tid$. 
To reason about such thread-observable views, \logic supports assertions of the form $\View x \tid = S$, stating that $\tid$ may read any value in the set $S$ for location $x$. That is, the coherence view of $\tid$ for $x$ consists of the writes whose values are those in $S$.

In the remainder of this article we enumerate the threads in our examples from left to right; \eg the left and right threads in \cref{fig:mp_proof} are identified as 1 and~2, respectively.
Moreover, we assume the registers of distinct threads have distinct names.
The precondition $P$ in \cref{fig:mp_proof} thus states that both threads may initially only read $0$ for both $x$ and $y$: $\for{\tid \!\in\! \{1, 2\}} [x]_\tid \!=\! [y]_\tid \!=\! \{0\}$. 


In the case of thread $1$, we can weaken $P$ (using the standard rule of consequence of Hoare logic -- see \rulename{Cons} in \cref{sec:logic}) to obtain $P_1$.
Upon executing $\store{x} 42$
\begin{enumerate*}[label=(\arabic*)]
	\item we weaken the resulting assertion by dropping the $a = 0$ conjunct; and 
	\item we update the observable view of thread $1$ on $x$ to reflect the new value of $x$: $\View x 1 = \{42\}$; that is, after executing $\store{x} 42$, the only value observable by thread $1$ for $x$ is $42$. 
\end{enumerate*}
Similarly, after executing $\store{y} 7$, we could assert $\View y 1 = \{7\}$; however, this is not necessary for establishing the final postcondition $Q$, and we thus simply weaken the postcondition to $\true$ ($P_3$).

Analogously, in the case of thread $2$ we weaken $P$ to obtain $Q_1$: $\View y 2 \!=\! \{0\}$ implies $\View y 2 \subseteq \{0, 7\}$ and $7 \in \valspred{y}{2} \imp \csetvalpred{y}{7}{x}{2} = \{42\}$. Note that $7 \in \valspred{y}{2} \imp \csetvalpred{y}{7}{x}{2} = \{42\}$ yields a vacuously true implication as $\View y 2 \!=\! \{0\}$ and thus $7 \not\in \View y 2$. 
The $\CView{y}{7}{x}{2}$ denotes a \emph{conditional view assertion}~\cite{DBLP:conf/ecoop/DalvandiDDW19}, capturing the essence of message passing by stating how reading a value on one location ($y$) affects the thread-observable view
on a different location ($x$).
More concretely, $\CView{y}{7}{x}{2} \!=\! \{42\}$ states that if thread $2$ executes a load on $y$ and reads value $7$, it subsequently may only observe value $42$ for $x$.  
This is indeed the essence of message passing in \ref{ex:mp}: once thread $2$ reads $7$ from $y$, it may only read $42$ for $x$ thereafter. 
As such, after executing the read instruction $\load{a} y$
\begin{enumerate*}[label=(\arabic*)]
	\item we apply the \rulename{LP_1} rule (in \cref{fig:proof-rules-pre-post}) which simply replaces $\View y 2$ with the local register $a$ in which the value of $y$ is read; and
	\item we replace the conditional assertion $\CView{y}{7}{x}{2} = \{42\}$ with the implication $a = 7 \imp \valspred{x}{2} = \{42\}$, stating that if the value read by thread $2$ for $y$ (in $a$) is $7$, then its observable view for $x$ is $\{42\}$. 
\end{enumerate*}
Similarly, upon executing $\load{b} x$ we simply apply \rulename{LP_1} to replace $\View x 2$ with the local register $b$ in which the value of $x$ is read.
%
%
Lastly, the final postcondition $Q$ is given by the conjunction of the thread-local postconditions ($P_3 \land Q_3$).

\vspace{2pt}
\noindent\textbf{\emph{Concurrent Reasoning and Stability}}
In our description of the \logic proof sketch in \cref{fig:mp_proof} thus far we focused on \emph{sequential} (per-thread) reasoning, ignoring how concurrent threads may affect the validity of assertions at each program point. 
Specifically, as in existing concurrent logics \cite{DBLP:journals/pacmpl/RaadLV20,DBLP:conf/ecoop/DalvandiDDW19,ogra,DBLP:journals/acta/OwickiG76}, we must ensure that the assertions at each program point are \emph{stable} under concurrent operations. 
For instance, to ensure that $P_1$ remains stable under the concurrent operation $\load{a}
 y$, we require that executing $\load{a} y$ on states satisfying the conjunction of $P_1$ and the precondition of $\load{a} y$ (\ie $Q_1$) not invalidate $P_1$, in that the resulting states continue to satisfy $P_1$; that is, $\assert{P_1 \land Q_1} \load{a} y \assert{P_1}$ holds. 
Similarly, we must ensure that $P_1$ is stable under $\load{b} x$, \ie $\assert{P_1 \land Q_2} \load{b} x \assert{P_1}$ holds. 
Analogously, we must establish the stability of $P_2$, $P_3$, $Q_1$, $Q_2$ and $Q_3$ under concurrent operations. 
In \cref{sec:logic} we present syntactic rules that simplify the task of checking stability obligations.
It is then straightforward to show that the assertions in \cref{fig:mp_proof} are stable.

\begin{figure}[t!]
\begin{center}
\begin{tabular}{@{} l @{\hspace{10pt}} | @{\hspace{10pt}} l @{}}
$
\begin{array}{@{} l @{}}
%
%
	\assert{\PView y = \{0\}} \\
	\qquad\store x 1; \annotate{\rulename{SP_1}} \\
	\assert{
		\tightshade{$\View x 1 = \{1\}$} 
		\land \PView y = \{0\}
	} \\
	\qquad\flC x;  \annotate{\rulename{FP_1}} \\
	\assert{\View x 1 = \{1\} \land \tightshade{$\PView x = \{1\}$} \land \PView y = \{0\}} \\
%
%
%
	\qquad\store y 1;  \annotate{\rulename{SP_1}} \\
	\assert{
		\View x 1 = \{1\} \land \PView x = \{1\} 
		\land \tightshade{$\View y 1 = \{1\}$}
	} \\
	\recassert{\PView y = \{1\} \imp \PView x = \{1\} }
\end{array}
$
& 
$
\begin{array}{@{} l @{}}
	\assert{\PView y = \{0\}} \\
	\qquad\store x 1; \annotate{\rulename{SP_1}} \\
	\assert{\tightshade{$\View x 1 = \{1\}$} \land \PView y = \{0\}} \\
	\qquad\foC x; \annotate{\rulename{OP_1}}\\
	\assert{\View x 1 {=} \{1\} \land \tightshade{$\AView x 1 {=} \{1\}$} \land \PView y {=} \{0\}} \\
	\qquad \sfenceC; \annotate{\rulename{SFP_1}}\\
	\assert{\View x 1 {=} \{1\} \land \tightshade{$\PView x {=} \{1\}$} \land \PView y {=} \{0\}} \\
	\qquad\store y 1; \annotate{\rulename{SP_1}}\\
	\assert{\View x 1 {=} \{1\} \land \PView x {=} \{1\} \land \tightshade{$\View y 1 {=} \{1\}$} } \\
	\recassert{\PView y = \{1\} \imp \PView x = \{1\} }
\end{array}
$
\end{tabular}\vspace{-5pt}
\end{center}
\hrule
\caption{Proof sketches of \cref{subfig:litmus_tests_w_fl} (left) and \cref{subfig:litmus_tests_w_fo_sf} (right)}
\label{fig:fl_fo_proof}
\end{figure}

\vspace{2pt}
\noindent\textbf{\emph{Reasoning about \fl Persistency}}
 To reason about the relaxed, buffered persistency of \pxes, Cho \etal
 \cite{DBLP:conf/pldi/ChoLRK21} introduce \emph{persistency views}, determining the
 possible \emph{persisted} values for each location; \ie the values of
 those writes that may have persisted to memory.  Note that the
 persistency view determines the possible values observable upon
 recovery from a crash.  By contrast, the (per-thread) coherence views
 determine the observable values during normal (non-crashing)
 executions, and have no bearing on the post-crash values.

Analogously, we extend \logic with assertions of the form $\PView x = S$, stating that the persistent view for $x$ includes writes whose values are given by $S$. 
To see this, consider the \logic proof sketch of \cref{subfig:litmus_tests_w_fl} in \cref{fig:fl_fo_proof} (left). 
Initially, $y$ holds $0$ in persistent memory: $\PView y = \{0\}$. 
(Note that the precondition could additionally include $\View x 1 = \View y 1 = \{0\} \land \PView x = \{0\}$ to denote that initially the thread may only observe $0$ for $x$ and $y$ and that $x$ holds $0$ in persistent memory; however, this is not needed for the proof and we thus forgo it.)

As before, after executing $\store{x} 1$, the observable value for $x$ is updated, as denoted by $\View x 1 = \{1\}$. 
Moreover, after executing $\flC x$, the persisted value for $x$, as denoted by $\PView x = \{1\}$, by committing (persisting) the observable value for $x$ ($\View x 1 = \{1\}$) to memory (see \rulename{FP_1} in \cref{fig:proof-rules-pre-post}).
Finally, after executing $\store{y} 1$, the observable value for $y$ is updated, as denoted by $\View y 1 = \{1\}$.

\vspace{2pt}
\noindent\textbf{\emph{Crash Invariants}}
Recall that {\color{reccol} \rectext: $y {=} 1 \Rightarrow x {=} 1$}
in \cref{subfig:litmus_tests_w_fl} denotes a \emph{crash invariant} in
that it describes the persistent memory upon recover from a crash at
\emph{any} program point.  This is because we have no control over
when a crash may occur.  To capture such invariants, in \logic we
write \emph{quadruples} of the form $\quadruple{P}{C}{Q}{I}$, where
$\triple P C Q$ denotes a Hoare triple and $I$ denotes the crash
invariant. If $C$ is a sequential program, $I$ must follow from
\emph{every} assertion (including $P$ and $Q$) in the proof. For
instance, in the proof outline of \cref{fig:fl_fo_proof} (left) all
four assertions imply the invariant
$\PView y = \{1\} \imp \PView x = \{1\} $. We discuss the meaning of
crash invariants for concurrent programs below.


\vspace{2pt}
\noindent\textbf{\emph{Reasoning about \fo Persistency}}
Recall that unlike \fl, \fo instructions (due to instruction reordering) may behave asynchronously and their effects may not take place immediately after execution. 
As such, unlike for $\flC x$, after executing $\foC x $ we cannot simply copy the observable view on $x$  to the persistent view on $x$. 

To capture the asynchronous nature of \fo, Cho \etal \cite{DBLP:conf/pldi/ChoLRK21} introduce yet another set of views, namely the \emph{thread-local asynchronous view}: the asynchronous view of thread $\tid$ on $x$ describes the values (writes) that will be persisted at a later time (asynchronously) by $\tid$ upon executing a barrier instruction. 
That is, 
\begin{enumerate*}
\item when thread $\tid$ executes $\foC x$, its asynchronous view of
  $x$ is advanced to at least its observable view of $x$; and 
\item when $\tid$ executes a barrier (\sfenceC, \mfenceC or RMW), then
  its persistent view for each location is advanced to at least its
  corresponding asynchronous view.
\end{enumerate*}
We model this in \logic{} by
\begin{enumerate*}
\item setting $\AView x \tid$ to be a subset of $\View x \tid$ when $\foC x$ is executed; and 
\item setting $\PView x$ to be a subset of $\AView x \tid$ (for each location $x$) when a barrier is executed. 
\end{enumerate*}



This is illustrated in the proof sketch of \cref{subfig:litmus_tests_w_fo_sf} in \cref{fig:fl_fo_proof} (right). 
In particular, unlike the proof sketch of \cref{subfig:litmus_tests_w_fl} in \cref{fig:fl_fo_proof} (left), after executing $\foC x$ we cannot simply copy the thread-observable view to the persistent view. 
Rather, we copy the thread-observable view $\View x 1$ to its asynchronous view and assert $\AView x 1 = \{1\}$; 
and upon executing the subsequent $\sfenceC$, we copy the thread-asynchronous view to the persistent view and assert $\PView x = \{1\}$.

\begin{figure}[t]
\begin{center}
\begin{tabular}{c @{\hspace{5pt}} || @{\hspace{5pt}} c}
\multicolumn{2}{c}
{ $\nameassert{P}{a = 0 \land\for{o \in \{x, y,z\}, \tid \in \{1, 2\}} \View{o}{\tid} = \PView{o} = \{0\}}$ }\\
$
\begin{array}{@{} l @{}}
	\nameassert{P_1}{\View{y}{2} = \{0\} \land \PView{z} = \{0\} \land  a = 0}\\
	\qquad\store x 1; \annotate{\rulename{SP_1}} \\
	\nameassert{P_2}{
		\View{y}{2} = \{0\} \land \PView{z} = \{0\} \land  a = 0  
		\land \tightshade{$\View{x}{1} = \{1\}$}
	} \\
	\qquad \flC x; \annotate{\rulename{FP_1}, \rulename{Cons}}\\
	\nameassert{P_3}{
		\tightshade{$\PView{x} = \{1\}$} 
	} \\
	\qquad\store{y}  1; \annotate{\rulename{SP_1}, \rulename{Cons}}\\
	\nameassert{P_4}{\PView x = \{1\}}
\end{array}
$   
& 
$
\begin{array}{@{} l @{}}
	\assert{ 
                \true
	} \\
	\qquad\load a y;  \\
	\assert{
		\true 
	} \\
	\qquad\ifC {a = 1} \\
		\qqquad \assert{a = 1} \\
		\qqqquad \store{z}  1 ; \\	
	\assert{
		\true 
	}	
\end{array}
$  \\
\multicolumn{2}{c}{
\hspace{90pt}$\nameassert Q {\PView x = \{1\}}$
} \\
\multicolumn{2}{c}{
\hspace{165pt}$\namerecassert I {\PView z = \{1\} \imp \PView x = \{1\}}$
} 
\end{tabular} \vspace{-5pt}
\end{center}
\hrule\vspace{5pt}
\caption{A \logic proof sketch of \cref{subfig:litmus_tests_fl_concurrent}}
\label{fig:mp_fl_proof}
\end{figure}

\vspace{2pt}
\noindent\textbf{\emph{Putting It All Together}} We next present a \logic proof sketch of \cref{subfig:litmus_tests_fl_concurrent} in \cref{fig:mp_fl_proof}.
The proof of the left thread is analogous to that in \cref{fig:fl_fo_proof} (left); the proof of the right thread is straightforward and applies standard reasoning principles. 
The final postcondition $Q$ is obtained by weakening the conjunction of per-thread postconditions. 

Note that the crash invariant $I$ follows from the assertions at each program point of thread 1 (\ie $P_1 \lor P_2 \lor P_3 \lor P_4 \imp I$). 
That is, the crash invariant must follow from the assertions at \emph{all} program points of \emph{some} thread (\eg thread 1 in \cref{fig:mp_fl_proof}). 
In the case of sequential programs (\eg in \cref{fig:fl_fo_proof}), this amounts to all program points (of the only executing thread). 
Intuitively, we must ensure that the crash invariant holds at every program point regardless of how the underlying state changes. As the assertions are stable under concurrent operations, it is thus sufficient to ensure that there exists some thread whose assertions at each program point imply the crash invariant.




\newcommand{\fina}{{\it fin}}
\newcommand{\ina}{{\it in}}

\section{The \logic Proof rules and Reasoning Principles} 
\label{sec:logic}
\label{sec:proof-rules-and-og}
We proceed with a description of our verification framework.
%
%
As with prior work~\cite{DBLP:conf/ecoop/DalvandiDDW19},
the view-based semantics for persistent TSO
\cite{DBLP:conf/pldi/ChoLRK21} allows us to use the standard
\owickigries{} rules
\cite{DBLP:series/txcs/AptBO09,DBLP:journals/acta/OwickiG76} for
compound statements. The main adjustment is the introduction of a new
specialised assertion language capable of expressing properties about
the different ``views'' described intuitively in \cref{sec:motivation}. 
As such, since view updates are highly non-deterministic, the
standard ``assignment axiom'' of Hoare Logic (and by extension
\owickigries{}) is no longer applicable. Moreover, unlike SC, reads in a
weak memory setting have a side-effect: their interaction with the
memory location being read causes the view of the executing thread to
advance. Therefore, we resort to a set of proof rules that describe
how views are modified and manipulated, as formalised by our
view-based assertions.

\subsection{The \logic Programming Language}
\label{subsec:formal-background}

\begin{figure}[t]
\small
\[
\begin{array}{@{} l @{}}
	\val,\vala \!\in\! \Val \eqdef \Nats
	\qquad
	\loc, \loca,\ldots \!\in\! \Loc 
	\qquad
  a,b,\ldots \!\in\! \Reg 
	\qquad
	\tid \!\in\! \TId \eqdef \Nats 
	\qquad
	i, j, k,\ldots \!\in\! \Label \\
\begin{array}{@{} r @{\hspace{2pt}} l @{}}
	\hat{a},\hat{b},\ldots \in \AuxVar   & 
  	\hspace{127pt}
  	\hat{e} \in \AuxExp ::= 
  	v \mid \hat{a} \mid \hat{e} {+} \hat{e} \mid \cdots 
	\\
	e \in \Exp ::= 
  	& v \mid a \mid e {+} e \mid \cdots 
  	\hspace{80pt}	
	B \in \BExp ::= \true \mid B \land B \mid \cdots \\
	\alpha \in \ASt ::= 
	& \skipst \mid \assign a e \mid \load{a}{\loc} \mid \store{\loc}{e} \\
	& \mid \cas{a}{\loc}{e}{e} \mid \sfence \mid \mfence \mid \flush{\loc} \mid \flushopt{\loc} \\
  	ls \in \LSt ::= 
  	&  \stepgoto{\alpha}{j} \mid \ifgoto{B}{j}{k} \mid \Aux{\stepgoto{\alpha}{j}}{\hat{a} := \hat{e}}  \\
	\prog \in \Prog \eqdef 
	& \TID \times \Label \to \LSt  	
	\hfill
	\pc \in \PC \eqdef \TID \to \Label
\end{array} 
\end{array}
\vspace{-5pt}
\]
\hrule\vspace{5pt}
\caption{The \logic domains and programming language
}
\label{fig:language}
\end{figure}
We present the programming language in \cref{fig:language}.
%
%
%
%
%
\emph{Atomic statements} (in $\ASt$) comprise $\skipst$, assignment, memory reads and writes, barrier instructions and explicit persists. 
Specifically, $\assign a e$ evaluates expression $e$ and returns it in (thread-local) register $a$; $\load{a}{\loc}$ reads from memory location $\loc$ and returns it in register $a$; and 
$\store{\loc}{e}$ writes the contents of register $a$ to location $\loc$.  
The \cas{$a$}{$\loc$}{$e_1$}{$e_2$} denotes `compare-and-set' on
location $\loc$, from the evaluated value of $e_1$ to the evaluated
value of $e_2$, and sets $a$ to $1$ if the CAS succeeds and to $0$,
otherwise.
Finally, $\mfence$ denotes a memory fence, $\sfence$ denotes a store
fence, and $\flush{\loc}$ and $\flushopt{\loc}$ denote explicit
persist instructions (see~\cref{sec:motivation}).

Formally, we model a program $\prog$ as a function mapping each pair $(\tid, i)$ of thread identifier and label to the \emph{labelled statement} (in $\LSt$) to be executed. 
A labelled statement may be 
\begin{enumerate*}
	\item a plain statement of the form $\stepgoto{\alpha}{j}$, comprising an atomic statement $\alpha$ to be executed and the label
$j$ of the next statement; 
	\item a conditional statement of the form $\ifgoto{B}{j}{k}$ to accommodate branching, which proceeds to label $j$ if $B$ holds and to $k$, otherwise; or
	\item a statement with an auxiliary update
          $\Aux{\stepgoto{\alpha}{j}}{\hat{a} := \hat{e}}$, which
          behaves as $\stepgoto{\alpha}{j}$, but in addition (in the
          same atomic step) updates the value of the auxiliary
          variable $\hat{a}$ with the auxiliary expression
          $\hat{e}$. 
\end{enumerate*}
It is well known that Owicki-Gries proofs require auxiliary variables
to record the history of executions to differentiate states that would
otherwise not be
distinguishable~\cite{DBLP:journals/acta/OwickiG76}. We show how
auxiliary variables are used in \logicname{} in the flush buffering
example (\cref{sec:flush-buffering}).

We track the control flow within each thread via the \emph{program
  counter function}, $\pc$, recording the program counter of each
thread.  We assume a designated label, $\initlabel \in \Label$,
representing the \emph{initial label}; \ie each thread begins
execution with $\pc(\tid) = \initlabel$.  Similarly,
$\finlabel \in \Label$ represents the \emph{final label}.  Moreover,
if $\pc(\tid) = i$ at the current execution step, then:
\begin{enumerate*}
\item when $\prog(\tid, i) \!=\! \stepgoto{\alpha}{j}$ or $\prog(\tid, i) \!=\!
  \Aux{\stepgoto{\alpha}{j}}{a := \hat{e}}$, then
  $\pc(\tid) \!=\! j$ at the next step;
 
\item when $\prog(\tid, i) \!=\! \ifgoto{B}{j}{k}$ at the current step, 
then if $B$ holds in the current state, then $\pc(\tid) \!=\! j$ at the next step; 
otherwise $\pc(\tid) \!=\! k$ at the next step.
\end{enumerate*}

\newcommand{\kwif}{{\bf if}\,}
\newcommand{\kwthen}{{\bf then}\,}
\newcommand{\kwelse}{{\bf else}\,}
\newcommand{\kwelseif}{{\bf elseif}\,}

\begin{example} \label{ex:prog4} The program in
  \cref{fig:mp_fl_proof}, assuming that the left thread has id $1$, is
  given as follows. The formalisation of the right thread is omitted,
  but is similar.\smallskip
  
  \hfill$\Pi \defeq 
  \left\{\begin{array}{@{}l@{~~~}l@{}l}
           (1, \iota) \mapsto \stepgoto{\store{x}{1}}{2}, 
           (1, 2) \mapsto \stepgoto{\flush{x}}{3}, \\
           (1, 3) \mapsto \stepgoto{\store{y}{1}}{\finlabel}, ... 
         \end{array} \right\}
  $\hfill{ }
  
\end{example}

\subsection{View-Based Expressions}
\label{subsec:assertions}
As with prior work on the RC11 model~\cite{igps}, 
we interpret \logic expressions directly over a view-based state. 
We use expressions tailored for the view-based \pxes model~\cite{DBLP:conf/pldi/ChoLRK21}, which allow us to express relationships between different system components, including the persistent memory.

Our expressions fall into one of four categories: 
\begin{enumerate*}
	\item {\em current view} expressions, which describe the current views
of different system components (\eg the persistent view); 
	\item {\em conditional view} expressions~\cite{DBLP:conf/ecoop/DalvandiDDW19}, which describe  a view on a location after reading a particular value on a {\em
  different} location; 
  	\item {\em last view} expressions, which hold if a component is viewing the last write to a location; and 
	\item {\em write-count} expressions, which describe the number of writes to
a location. 
\end{enumerate*}


Our current view expressions comprise $\View{x}{\tid}$, $\PView{x}$ and $\AView{x}{\tid}$, as described below; as shown in \cref{sec:motivation}, each of these expressions describes a {\em set} of possible values.
\begin{description}
\item [$\View{x}{\tid}$] denotes the {\em coherence view} of thread $\tid$: the set
  of values $\tid$ may read for $x$.
\item [$\PView{x}$] denotes the {\em persistent memory view}: 
  the set of values that $x$ may hold in (persistent) memory. 

\item [$\AView{x}{\tid}$] denotes the {\em asynchronous memory view}
  of thread $\tid$: the set of values that can be persisted after a
  barrier instruction (\sfence/\mfence/RMW) is executed by $\tid$ (see
  rule {\sf OP} in \cref{fig:proof-rules-pre-post}). Asynchronous
  views are updated after executing a $\fo$; however, unlike
  persistent memory views, the values in asynchronous views are not
  guaranteed to be persisted until a subsequent barrier is executed by
  the same thread.
\end{description}

Conditional view expressions are of the form $\CView{x}{v}{y}{\tid}$, as described below. 
As discussed in \cref{sec:motivation}, conditional expressions capture the crux of message passing. 
\begin{description}
\item [$\CView{x}{v}{y}{\tid}$] returns a set of values that $\tid$
  may read for $y$ after it reads value $v$ for $x$. 
  In particular, if $\CView{x}{v}{y}{\tid} = S$ holds for
  some set $S$ and $\tid$ executes $\load{a}{x}$, then in the state
  immediately after the load, if $a = v$, then $\View{y}{\tid} \subseteq S$ (see
  $\rulename{LP_2}$ in \cref{fig:proof-rules-pre-post}).
\end{description}

Last-view expressions~(\cf~\cite{DBLP:conf/ppopp/DohertyDWD19}) are
boolean-valued and hold if a particular component is synchronised (\ie observes the latest value) on the given location. Such expressions provide
determinism guarantees on $\loadC$ and $\fl$. 
For instance if the view of $\tid$ is the last write on $x$, then a read from $x$ by $\tid$ will load this last value. Last-view expressions comprise $\LastR{x}{\tid}$ and $\LastF{x}{\tid}$:
\begin{description}
\item [$\LastR{x}{\tid}$] holds iff $\tid$ is currently viewing the
  {\em last} write to $x$. Thus, for example, if $\LastR{x}{\tid}$
  holds, then a \ld from $x$ by $\tid$ reads the last write to
  $x$. Note that unlike architectural operational models \cite{tso},
  in the view model \cite{DBLP:conf/pldi/ChoLRK21}, writes are visible
  to all threads as soon as they occur.
\item [$\LastF{x}{\tid}$] holds iff a \flush of $x$ by $\tid$ is
  guaranteed to flush the {\em last} write to $x$ to persistent
  memory.
\end{description}

Lastly, write-count expressions are of the form $\oc{x}{v}$, as
described below. Such assertions are useful for inferring view
expressions from known facts about the number of writes in the system
with a particular value (see \cref{fig:po-ep2}).
\begin{description}
\item [$\oc{x}{v}$] returns the number of writes to $x$ with
  value $v$. If $\oc{x}{v}$ holds and $\tid$ writes to
  $y \neq x$, or writes a value $u \neq v$, then $\oc{x}{v}$ continues to hold  afterwards.
\end{description}

\subsection{\owickigries{} Reasoning}
\label{subsec:owicki-gries-reas}

We present the \logic proof system, as an extension
of Hoare Logic with \owickigries{} reasoning to account for
concurrency. The main differences are that
\begin{enumerate*}
	\item our program annotations contain view-based assertions that allow reasoning about weak and persistent memory behaviours; and 
	\item we define a crash invariant to describe the recoverable state of the program after a
crash. 
\end{enumerate*}
We proceed by first defining proof outlines, then providing syntactic rules for proving their validity. Our proof rules are \emph{syntactic}, and thus can be understood and used without having to understand the details of the underlying \pxes model. 

We let $\Assertion$ be the set of \emph{assertions} (\ie predicates over
\pxes states) that use view-based expressions (\cref{subsec:assertions}). 
A \emph{crash invariant}, $\inv \in \Inv \subset \Assertion$, is defined over persistent views only, \ie it only comprises the persistent view expressions of the form $\PView x$.
We model program annotations via an \emph{annotation function}, $\ann \in \Ann = \TID \times \Label \to \Assertion$, associating each program point $(\tid, i)$ with its associated assertion.
A {\em proof outline} is a tuple $(\ina, \ann, \inv, \fina)$, where $\ina, \fina \in \Assertion$ are
the initial and final assertions.

\begin{example} \label{ex:po4} The annotation of the
  proof in \cref{fig:mp_fl_proof} is given by $\ann$, with the mappings of thread $1$ as shown below; the mappings of thread $2$ are similar.

  \smallskip\hfill$\ann \defeq 
  \left\{\begin{array}{@{}l@{~~~}l@{}l}
           (1, \iota) \mapsto P_1,
           (1, 2) \mapsto P_2, 
           (1, 3) \mapsto P_3, 
           (1, \finlabel) \mapsto P_4, \dots 
         \end{array} \right\}
       $\hfill{}

       \smallskip
       \noindent Additionally, we have
       $in \defeq a = 0 \land\for{o \in \{x,y,z\}, \tid \in \{1, 2\}}
       \View{o}{\tid} = \PView{o} = \{0\}$, $\fina
       \defeq 
       \PView x = \{1\}$ and
       $\inv \defeq \PView z = \{1\} \imp \PView x = \{1\}$.
\end{example}

\begin{definition}[Valid proof outline]{\rm
\label{def:outline}
A proof outline $(\ina, \ann, \inv, \fina)$
is {\em valid} for a program $\prog$ iff the following hold: 
  \begin{description}
  \item [\sf Initialisation.] For all $\tid \in \TID$,
    $\ina \imp \ann(\tid, \initlabel)$.
    \item [\sf Finalisation.] 
    $(\bigwedge_{\tid \in \TID}\ \ann(\tid, \finlabel)) \imp \fina$. 
  \item [\sf Local correctness.] For all $\tid \in \TID$ and $i \in \Label$, either:
    \begin{itemize}
    \item $\prog(\tid, i) = \stepgoto{\alpha}{j}$ and
      ${\assert{\ann(\tid, i)}}\ \alpha\ {\assert{\ann(\tid, j)}}$; or 
    \item $\prog(\tid, i) = \ifgoto{B}{j}{k}$ and both
      $\ann(\tid, i) \wedge B \imp \ann(\tid,j)$ and
      $\ann(\tid, i) \wedge \neg B \imp \ann(\tid,k)$ hold; or
    \item $\prog(\tid, i) = \Aux{\stepgoto{\alpha}{j}}{\hat{a} := \hat{e}}$ and
      ${\assert{\ann(\tid, i)}}\ \alpha\ {\assert{\ann(\tid, j)[\hat{e} / \hat{a}]}}$.
    \end{itemize}
  
  \item [\sf Stability.]
    For all $\tid_1, \tid_2 \in \TID$ such that $\tid_1 \neq \tid_2$ and
    $i_1, i_2 \in \Label$:

    \begin{itemize}
    \item if $\prog(\tid_1, i_1) = \stepgoto{\alpha}{j}$, then $\assert{\ann(\tid_2, i_2) \wedge \ann(\tid_1, i_1)}\ \alpha\ \assert{\ann(\tid_2, i_2)}$; 
    \item if $\prog(\tid_1, i_1) = \Aux{\stepgoto{\alpha}{j}}{\hat{a} := \hat{e}}$, then
      
      $\assert{\ann(\tid_2, i_2) \wedge \ann(\tid_1, i_1)}\ \alpha\ \assert{\ann(\tid_2, i_2)[\hat{e}/\hat{a}]}$. 
    \end{itemize}

          

  \item [\sf Persistence.]  There exists $\tid \in \TID$ such that for all
    $i \in \Label$, $\ann(\tid, i) \imp \inv$.
  \end{description}

}\end{definition} Intuitively, {\sf Initialisation}
(\resp\textsf{Finalisation}) ensures that the initial (\resp final)
assertion of each thread holds at the beginning (\resp end); {\sf
  Local correctness} establishes annotation validity for each thread;
{\sf Stability} ensures that each (local) thread annotation is {\em
  interference-free} under the execution of other
threads~\cite{DBLP:journals/acta/OwickiG76}; and {\sf Persistence}
ensures that the crash invariant holds at every program point for some
thread.


\begin{example}
  \label{ex:ob4}
  Given the program in \cref{ex:prog4} and its annotation in
  \cref{ex:po4}, both {\sf Initialisation} and {\sf Finalisation}
  clearly hold. Moreover, {\sf Persistence} holds for thread $1$.  
  For {\sf Local correctness} of thread $1$, we
  must prove \eqref{eq:lc1-1}--\eqref{eq:lc1-3} below
  ; 
  {\sf Local correctness} of thread $2$ is
  similar.
  \begin{eqnarray}
    \label{eq:lc1-1}
    \assert{
      P_1
    }
    &
      \store{x}{1} & \assert{
        P_2 
    }\\
    \label{eq:lc1-2}
    \assert{
      P_2 
    } & \flush{x} & \assert{P_3}
    \\
    \label{eq:lc1-3}
    \assert{P_3 
    } & \store{y}{1} & \assert {P_4}
  \end{eqnarray}
  For {\sf Stability} of $P$ (the precondition of $\store{x}{1}$ in
  thread 1) against thread 2 we must prove:
    \begin{eqnarray}
    \label{eq:st1-1}
      \assert{
      P_1 
    }
    &
      \load{a}{y} & \assert{
        P_1
    }\\
    \label{eq:st1-2}
    \assert{P_1 \wedge a =1} & \store{z}{1} & \assert {P_1}
    \end{eqnarray}
    {\sf Stability} of other assertions (\ie, $P_2$--$P_4$)
    is similar.  We prove \eqref{eq:lc1-1}--\eqref{eq:st1-2} in
    \cref{subsec:proof-rules}.
\end{example}

\subsection{\logic Proof rules}
\label{subsec:proof-rules}

One of the main benefits of \logic is the ability to perform
proofs at a high level of abstraction.  In this section, we provide
the set of proof rules that we use. The annotation within a proof
outline is, in essence, an invariant mapping each program location to
an assertion that holds at the program location. Thus, we prove local
correctness by checking that each atomic step of a thread establishes
the assertions in that thread. Similarly, we check stability by
checking each assertion in one thread against each atomic step of the
other threads. To enable proof abstraction, we introduce a set of
proof rules that describe the interaction between the 
assertions from \cref{subsec:assertions} and the atomic program
steps. We will use the standard decomposition rules from Hoare Logic
to reduce proof outlines and enable our rules over atomic steps to be
applied.

\smallskip
\noindent\textbf{\emph{Standard Decomposition Rules}} The standard
decomposition rules we use are given in \cref{fig:decomp}, which
allow one to weaken preconditions and strengthen postconditions, and
decompose conjunctions and disjunctions.
\begin{figure}[t]
  \centering
  $\inference[\rulename{Cons}]{\!\!\!\!\!\!\!P' \imp P \quad Q \imp Q'\!\!\!\!\!\! \\ \{P\}~\prog~\{Q\}}{\!\!\!\!\!\{P'\} ~\prog~ \{Q'\}\!\!\!\!\!\!\!}$
  \quad
  $\inference[\rulename{Conj}]{\{P_1\}~\prog~\{Q_1\}\\ \{P_2\}~\prog~\{Q_2\} }{\!\!\!\!\{P_1 \wedge P_2 \} ~\prog~ \{ Q_1 \wedge Q_2\}\!\!\!\!}$
  \quad
  $\inference[\rulename{Disj}]{\{P_1\}~\prog~\{Q_1\}\\ \{P_2\}~\prog~\{Q_2\} }{\!\!\!\!\{P_1 \vee P_2 \} ~\prog~ \{ Q_1 \vee Q_2\}\!\!\!\!}$
         \vspace{-5pt}
         \caption{Standard decomposition rules of \logic}
  \label{fig:decomp}
\end{figure}


\smallskip
\noindent\textbf{\emph{Rules for Atomic Statements and View-Based Assertions}}
Weak and persistent memory models (\eg Px86) are inherently
non-deterministic. Moreover in contrast to sequential consistent, in
view-based operational semantics (such as \pxes) instructions such as
 $\load{a}{e}$ have may a side-effect since they may update the view
of the thread performing the $\loadC$
(\cf~\cite{DBLP:conf/ecoop/DalvandiDDW19}). Therefore, unlike Hoare
Logic, which contains a single rule for assignment, we have a set of
rules for atomic statements, describing their interaction with
view-based assertions. Each of the rules in this section has been
proved sound with respect to the view-based semantics in
Isabelle/HOL. 

\begin{figure}[t]
  \centering
  \begin{array}[t]{r@{~}c@{~}l@{~}|@{~}l@{~}|@{~}l}
    \textrm{Precondition} & \textrm{Statement} & \textrm{Postcondition} & \textrm{Const.} & \textrm{Ref.}  \\
    \hline  
    \assert{\View{x}{\tid} =  S} & \multirow{3}{*}{$\load{a}{x}$} & \assert{a \in S \wedge \View{x}{\tid}\subseteq S} & & \rulename{LP_1}
    \\
    \assert{u \in \View{x}{\tid}  \imp \CView{x}{u}{y}{\tid} = S} &  & \assert{a = u \imp \View{y}{\tid} \subseteq S} && \rulename{LP_2}
    \\
    \assert{\begin{array}[c]{@{}l@{}}
              \oc{x}{u}=1 \wedge {} 
              \LastR{x}{\tid'} \wedge \View{x}{\tid'} = \{u\}
      \end{array}} & 
& \assert{a=u \imp \View{x}{\tid} = \{u\}} & & \rulename{LP_3}
    \\
    \hline    
    \assert{true} & \multirow{7}{*}{$\store{x}{v}$ }  & \assert{\valspred{x}{\tid} = \{v\}} 
       & & \rulename{SP_1}
     \\
     \assert{\valspred{x}{\tid'}  = S} & 
     & \assert{\valspred{x}{\tid'}  = S \cup \{v\}} & \tid \neq \tid' & \rulename{SP_2}
     \\
     \assert{\AView{x}{\tid'}  = S} & & \assert{\AView{x}{\tid'}  = S \cup \{v\}} & & \rulename{SP_3}
     \\
     \assert{\PView{x}  = S} &   & \assert{\PView{x}  = S \cup \{v\}} & & \rulename{SP_4}
     \\
     \assert{\View{y}{\tid} = S \wedge v \notin \View{x}{\tid'}}
                       & 
     & \assert{\CView{x}{v}{y}{\tid'} \subseteq S}
                                      &  \tid \neq \tid' & \rulename{SP_5}
     \\
     \assert{true} &   & \assert{\LastR{x}{\tid} \wedge \LastF{x}{\tid} 
                         }  & & \rulename{SP_6}
     \\
    \assert{\oc{x}{v}= n} & & \assert{\oc{x}{v}= n + 1} & & \rulename{SP_7}
    \\
    \hline
    \assert{\View{x}{\tid} = S}  & \multirow{3}{*}{$\flush{x}$ } & \assert{\PView{x}  \subseteq S \land \AView{x}{\tid}  \subseteq S} & & \rulename{FP_1}
    \\
    \assert{\PView{x}  = S} &  & \assert{\PView{x} \subseteq S} & & \rulename{FP_2}
    \\
    \assert{\LastR{x}{\tid'} \wedge 
    \valspred{x}{\tid'} = \{u\} 
    \wedge \LastF{x}{\tid}
    }
                          &  & \assert{\PView{x} = \{u\}}  && \rulename{FP_3}    \\
    \hline
    \assert{\valspred{x}{\tid} = S \lor \AView{x}{\tid}  = S} & \flushopt{x} & \assert{\AView{x}{\tid}  \subseteq S} & & \rulename{OP}
 \\
    \hline
    \assert{\AView{x}{\tid} = S \lor \PView{x} = S} &    \sfence
    &
      \assert{\PView{x}   \subseteq S}  & & \rulename{SFP} 
  \end{array}
  \caption{Selected proof rules for atomic statements executed by
    thread $\tid$ } 
  \label{fig:proof-rules-pre-post}
\end{figure}
A selection of these rules for the atomic statements is
given in \cref{fig:proof-rules-pre-post}, where the statement is
assumed to be executed by thread $\tid$.  The first column contains
the pre/post condition triple, the second any additional constraints
and the third, labels that we use to refer to the rules in our
descriptions below. Unless explicitly mentioned as a constraint, we do
not assume that threads, locations and values are distinct; \eg rule
$\rulename{LP_3}$ (referring to $\tid$ and $\tid'$) holds regardless
of whether $\tid = \tid'$ or not. 

The rules in \cref{fig:proof-rules-pre-post} 
provide high-level
insights into the low-level semantics of \pxes without having to
understand the operational details. The $\rulename{LP_i}$ rules are
for statement $\load{a}{x}$. Rule $\rulename{LP_1}$ states that if
$\tid$'s view of $x$ is the set of values $S$, then in the post state
$a$ is an element of $S$ and moreover $\tid$'s view of $x$ is a subset
of $S$ (since $\tid$'s view may have shifted).  By $\rulename{LP_2}$,
provided the conditional view of $\tid$ on $y$ (with condition
$x = u$) is $S$, if the load returns value $u$, then the view of
$\tid$ is shifted so that $\View{y}{\tid} \subseteq S$. We only have
$\View{y}{\tid} \subseteq S$ in the postcondition because there may be
multiple writes to $x$ with value $u$; reading $x$ read may shift the
view to the latter write, thus {\em reducing} the set of values that
$\tid$ can read for $y$. $\rulename{LP_3}$ describes conditions for a
deterministic load by thread $\tid$. The precondition assumes that
there is only one write to $x$ with value $u$, that {\em some} thread
$\tid'$ sees the last write to $x$ with value $u$. Then, if $\tid$
reads $u$, its view of $x$ is also constrained to just the set
containing $u$.

The store rules, $\rulename{SP_i}$, reflect that fact that a new write
modifies the views of the other threads as well as the persistent
memory and asynchronous views. The first four rules describe the
interaction of a $\storeC$ by thread $\tid$ with current view
assertions. By $\rulename{ SP_1}$, the $\storeC$ ensures that the
current view of $\tid$ is solely the value $v$ written by $\tid$. This
is because in $\pxes$, new writes are introduced by the executing
thread, $\tid$, with a maximal timestamp (see {\sc store} rule in
\cref{fig:full:viewpintel}), and $\tid$'s view is updated to this new
write. $\rulename{SP_2}$, $\rulename{SP_3}$ and $\rulename{SP_4}$ are
similar, and assuming that the view (of another thread, persistent
memory and asynchronous view, respectively) in the pre-state is $S$,
shows that the view in the post state is $S \cup \{v\}$. Rule
$\rulename{SP_5}$ allows one to {\em introduce} a conditional
observation assertion $\CView{x}{v}{y}{\tid'}$ where
$\tid' \neq \tid$. The pre-state of $\rulename{SP_5}$ assumes that
$\tid$'s view of $y$ is the set $S$, and that $\tid'$ cannot view
value $v$ for $y$. Rule $\rulename{SP_6}$ introduces last-view
assertions for $\tid$ after $\tid$ performs a write to $x$, and
finally $\rulename{SP_7}$ states that the number of writes to $x$ with
value $v$ increases by $1$ after executing $\store{x}{v}$.

Rules $\rulename{FP_i}$ describe the effect of $\flush{x}$ on the
state. $\rulename{FP_1}$ states that, provided that the current view of
$\tid$ for $x$ is the set of values $S$, after executing $\flush{x}$,
we are guaranteed that both the persistent view and asynchronous view
of $\tid$ for $x$ are subsets of $S$. We obtain a subset in the post
state since the \pxes semantics potentially moves the persistent and
asynchronous views forward. Similarly, by $\rulename{FP_2}$ if the current
persistent view of $x$ is $S$, then after executing $\flush{x}$ the
persistent view will be a subset of $S$. Finally, $\rulename{FP_2}$
provides a mechanism for establishing a deterministic persistent view
$u$ for $x$. The precondition assumes that {\em some} thread's view of
$x$ is the last write with value $u$ and that $\tid$'s view is such
that the flush is guaranted to flush to this last write to $x$.

Rule $\rulename{OP}$ describes how the asynchronous view of $\tid$ in the
postcondition of $\flushopt{x}$ is related to the current view of
$\tid$ and the asynchronous view in the precondition. Finally, rule
$\rulename{SFP}$ describes the relationship between the persistent view in
the postcondition and the asynchronous view and persistent view in the
precondition for an $\sfenceC$ instruction.

Our Isabelle/HOL development contains further rules for the other
instructions, including $\mfence$ and $\casC$, which we omit here
for space reasons. In addition, we prove the stability of several
assertions (see \cref{fig:proof-rules-stable} for a selection). An
assertion $P$ is {\em stable} over a statement $\alpha$ executed by
$\tid$ iff $\{P\}~\alpha~\{P\}$ holds.




\begin{figure}[t]
  \centering
  
  \fontsize{8.7}{10}\selectfont
  \begin{array}[t]{@{}l@{~}||@{~}l@{}}
    \begin{array}[t]{l@{~}|@{\,}l@{\,}|@{\,}l@{\,}|@{\,}l}
    \textrm{Statement} & \textrm{Stable Assert.}  & \textrm{Const.} & \textrm{Ref.}  \\
      \hline
      \multirow{5}{*}{$\load{a}{x}$ } 
    & \assert{\View{y}{\tid'} = S} & \tid \neq \tid'& \rulename{LS_1} 
    \\
    & \assert{\PView{y} = S}  & & \rulename{LS_2} 
    \\
    & \assert{\AView{y}{\tid'} = S}  && \rulename{LS_3} 
      \\
    & \assert{a = k} && \rulename{LS_4}
    \\
    & \assert{\LastR{y}{\tid'}} & & \rulename{LS_5} 
      \\
      \hline
            \multirow{5}{*}{$\flush{x}$ } 
    & \assert{\View{y}{\tid'} = S} & & \rulename{FS_1} 
    \\
    & \assert{\PView{y} = S} & x \neq y &  \rulename{FS_2} 
    \\
    & \assert{\LastR{y}{\tid'}} & & \rulename{FS_3} 
    \\
    & \assert{\LastF{y}{\tid'}}  &  &  \rulename{FS_4} 
    \\
    & \assert{\oc{y}{v}= n} && \rulename{FS_5} 
    \\
      \hline
                \multirow{2}{*}{$\sfence$ } 
 & \assert{\View{x}{\tid'} = S} & & \rulename{SFS_1} 
    \\
    & \assert{\oc{x}{v}= n} & & \rulename{SFS_2}
    \end{array}
                                                                    &
    \begin{array}[t]{l@{\,}|@{\,}l@{\,}|@{\,}l@{\,}|@{\,}l}
    \textrm{Statement} & \textrm{Stable Assert.}  & \textrm{Const.} & \textrm{Ref.}  \\
    \hline
                    \multirow{7}{*}{$\store{x}{v}$ } 
      &
          \assert{\valspred{y}{\tid'} = S}  & x \neq y& \rulename{WS_1} 
    \\
    & \assert{\PView{y}  = S}  & x \neq y & \rulename{WS_2} 
    \\
    & \assert{\AView{y}{\tid'}  = S}  & x \neq y& \rulename{WS_3} 
      \\
    & \assert{a = k}  & & \rulename{WS_4} 
      \\
    & \assert{\LastR{y}{\tid'}}  & x \neq y& \rulename{WS_5} 
      \\
    & \assert{\LastF{y}{\tid'}}  & x \neq y& \rulename{WS_6} 
      \\
    & \assert{\oc{y}{v'}= n}  &
                                \begin{array}[t]{@{}l@{}}
                                  x \neq y \lor{} \\
                                  v \neq v'
                                \end{array}
                       & \rulename{WS_7}
      \\
      \hline
                          \multirow{3}{*}{$\flushopt{x}$ } 
    & \assert{\View{y}{\tid'} = S} & & \rulename{OS_1} 
    \\
    & \assert{\PView{y} = S} & & \rulename{OS_2} 
    \\
    & \assert{\oc{y}{v}= n} & &\rulename{OS_3} 
    \end{array}
  \end{array}
  \caption{Selection of stable assertions for atomic statements
    executed by thread $\tid$  }
  \label{fig:proof-rules-stable}
\end{figure}

\smallskip
\noindent\textbf{\emph{Well-formedness}} The final major aspect of our framework is
a well-formedness condition that describes the set of reacahble states
in the \pxes semantics. The condition is expressed as an invariant of
the semantics: it holds initially, and is stable under every possible
transition of \pxes. In fact, the rules in
Figs.~\ref{fig:proof-rules-pre-post} and~\ref{fig:proof-rules-stable}
are proved with respect to this well-formedness condition. 

The majority of the well-formedness constraints are straightforward,
\eg describing the relationship between the views of different
components.  The most important component of the well-formedness
condition is a non-emptiness condition on views, which states that
$ \View{x}{\tid} \neq \emptyset \wedge \PView{x} \neq \emptyset \wedge
\AView{x}{\tid} \neq \emptyset $. For instance, a consequence of this condition is
that, in combination with \rulename{LP_1}, we have:
\begin{align}
  \label{eq:lp1-cor}
  {\assert{  \valspred{y}{\tid} = \{v\}}} ~\load{a}{x}~{\assert{\valspred{y}{\tid} = \{v\}}}
\end{align}


\smallskip
\noindent\textbf{\emph{Worked Example}}
 We now return to the proof obligations from \cref{ex:ob4} and
 demonstrate how they can be discharged using the proof rules
 described above.
  For {\sf Local correctness}, condition \eqref{eq:lc1-1} holds by
  {\sf Conj} (from \cref{fig:decomp}) together with stability rules
  $\rulename{WS_1}$, $\rulename{WS_2}$ and $\rulename{WS_4}$ (from
  \cref{fig:proof-rules-stable}) which establish the first three
  conjunctions in the postcondition, and $\rulename{SP_1}$ from
  \cref{fig:proof-rules-pre-post}, which establishes the final
  conjunction. Condition \eqref{eq:lc1-2} holds by $\rulename{FP_1}$ in
  \cref{fig:proof-rules-pre-post} together with {\sf Cons} (from
  \cref{fig:decomp}). Finally, condition \eqref{eq:lc1-3} holds by
  $\rulename{WS_2}$ (from \cref{fig:proof-rules-stable}).

  Both the {\sf Stability} conditions \eqref{eq:st1-1} and
  \eqref{eq:st1-2} from \cref{ex:ob4} hold by the stability rules in
  \cref{fig:proof-rules-stable} together with {\sf Cons} and {\sf
    Conj} (from \cref{fig:decomp}). In particular, for
  \eqref{eq:st1-1}, we use rules $\rulename{LS_1}$, $\rulename{LS_2}$
  and $\rulename{LS_4}$, and for \eqref{eq:st1-2}, we use
  $\rulename{WS_1}$, $\rulename{WS_2}$ and $\rulename{WS_4}$.

 \section{Formal definitions of assertions}







 \begin{verbatim}


(* Auxiliary definition for the set of conditional observation *)
definition " compvrnew ts ti ind l ≡ 
 let  nview =  if( ind \neq (coh ts ti) l )
                   then ind 
                else vrnew ts ti
in  max nview (vrnew ts ti)"


(* Auxiliary definition for the set of conditional observation *)
definition " cond_ts ts ti l v ≡ {compvrnew ts ti t l   | t. t ∈ OTS ts ti l \wedge v = compval  ((memory ts)!t)}"


(* Set of conditionally observable timestamps: COBTS *)
definition " COBTS x y v ti ts ≡  ⋃ { OTSF ts ti y nview | nview.   nview  ∈ cond_ts ts ti x v} "


 
 definition mapval
  where "mapval  B li = (\lambda x. compval( (li)!x) ) ` B"


abbreviation COBVab :: "Loc  \Rightarrow Val  \Rightarrow Loc  \Rightarrow TId \Rightarrow TState \Rightarrow Val set" ("⟨_  _⟩[_]⇩_ _ " [100, 100, 100, 100, 100])
  where "⟨x  v⟩[y]⇩t  \sigma ≡  mapval (COBTS x y v t \sigma) (memory \sigma) "


(* Set of all indexes in memory that point to messages with location addr *)
definition " last_entry_set ts addr ≡  { i  | i. 0 \leqi \wedge i < length(memory ts) \wedge comploc ((memory ts)!i) addr = addr}"


(* Index in memory that point to last  message with location addr *)
definition " last_entry  ts addr  ≡ Max ( last_entry_set ts addr )"


(*new*)
abbreviation OTS_local_max ::" Loc \Rightarrow TId \Rightarrow TState \Rightarrow bool" ("⌈_⌉⇩_ _" [100,100,100])
  where
 "⌈l⌉⇩t \sigma ≡  (OTS  \sigma t l) = { last_entry  \sigma  l  }  "


abbreviation flush_order :: " Loc \Rightarrow TId \Rightarrow  TState \Rightarrow bool" ( "⌈FLUSH _ ⌉⇩_ _ "  [100,100,100])
 where
 "⌈FLUSH l⌉⇩t \sigma ≡   last_entry  \sigma  l \leq max  (max ( maxcoh \sigma t)  (vpcommit \sigma t l))  (maxvp \sigma l) "



abbreviation mfence_order :: " Loc \Rightarrow TId \Rightarrow  TState \Rightarrow bool" ( "⌈MFENCE _ ⌉⇩_ _ "  [100,100,100])
 where
 "⌈MFENCE l⌉⇩t \sigma ≡   last_entry  \sigma  l \leq max (vrnew \sigma t)  (maxcoh \sigma t) "



definition " oc_set ts  x v ≡  {t | t.  0\leq t \wedge t < length(memory ts) \wedge x= comploc ((memory ts)!t) x \wedge compval  ((memory ts)!t) = v}"


abbreviation count_oc :: " Loc \Rightarrow Val \Rightarrow TState \Rightarrow nat " ( "⦇_,_⦈  _" [100,100,100])
  where
" ⦇x,v⦈  \sigma  ≡ card  (oc_set \sigma  x v) " 



\end{verbatim}


\section{Examples} 
\label{sec:example-proofs}

In this section we present a selection of programs that we have
verified in Isabelle/HOL. These examples highlight specific aspects of
Px86, in particular, the interaction between $\fo$ and $\sfenceC$, as
well as aspects of our view-based assertion language that simplifies
verification.


\vspace{5pt}\noindent\textbf{\emph{Optimised Message Passing}} We
start by considering a variant of
\cref{subfig:litmus_tests_fl_concurrent}, which contains two
optimisations. First, we notice that flushing of the write to $x$ in
thread $1$ can be moved to thread $2$ since the write to $z$ is
guarded by whether or not thread $2$ reads the flag $y$. Second, it is
possible to replace the $\flushC$ by a more optimised $\fo$ followed
by an $\sfenceC$. We confirm correctness of these optimisations via the
proof outline in \cref{fig:po-omp}. The optimised message passing in \cref{fig:po-omp} ensures
the same persistent invariant as
\cref{subfig:litmus_tests_fl_concurrent}. However, the way in which
this is established differs. In particular, in
\cref{subfig:litmus_tests_fl_concurrent}, the persistent invariant
holds due to thread 1, whereas in \cref{fig:po-omp} it holds due to
thread 2.

\begin{figure}[!t]
  \centering
\begin{tabular}{c||c}
\multicolumn{2}{c}
{ $\assert{\forall o \in \{x,y,z\}, \tid \in \{1,2\} \ldotp \View{o}{\tid} =  \PView{o} = \AView{o}{\tid} = \{0\}}$ }\\
$\begin{array}{l}
\assert{\View{y}{2}=\{0\}} \\
\store{x}{1} ;\\
\assert{
   \begin{array}{l}
     \View{y}{2}=\{0\} \wedge {}\\
     \View{x}{1} = \{1\}
   \end{array}
} \\
\store{y}{1};  \\
\assert{\true}\\
\end{array}$ 
& 
$\begin{array}{l}
\assert{
   \begin{array}{l}
     (1 \in \View{y}{2} \imp \csetvalpred{y}{1}{x}{2} = \{1\} )\wedge {} 
     \View{y}{2} \subseteq \{0, 1\} \wedge  \PView{z} = \{0\}
   \end{array}
} \\
\load{a}{y};  \\
\assert{ (a = 1 \imp \View{x}{2} = \{1\}) \wedge \PView{z} = \{0\}}\\
 \ifC {a \neq 0} \\
\qquad \assert{ \View{x}{2} = \{1\} \wedge \PView{z} = \{0\}}\\
\qquad     \flushopt{x}; \\  
\qquad \assert{ \AView{x}{2} = \{1\} \wedge \PView{z} = \{0\}}\\
\qquad     \sfence;      \\
\qquad \assert{ \PView{x} = \{1\}}\\
\qquad    \store{z}{1};   \\
\assert{\PView{z} = \{0\} \lor \PView{x} = \{1\}}
\end{array}$
\\
\multicolumn{2}{c}
{ $\assert{
\PView{z} = \{0\} \lor \PView{x} = \{1\} }$ \quad \qquad \qquad \qquad }
  \\
\multicolumn{2}{c}
{$\recassert{\PView{z}=\{1\} \imp  \PView{x}=\{1\}}$} \qquad \qquad \qquad 
\end{tabular}
\vspace{-5pt}
  
  \caption{Proof outline for optimised message passing}
  \label{fig:po-omp}
\end{figure}


With respect to the persistent invariant, the most important sequence
of steps takes place in thread 2 if it reads $1$ for $y$. Note that by
the conditional view assertion in the precondition of $\load{a}{y}$,
thread $2$ is guaranteed to read $2$ for $x$ after reading $1$ for
$y$. Thus, if the test of $\kwc{if}$ statement succeeds, then thread
$2$ must see $1$ for $x$. This view is translated into an asynchronous
view after the $\fo$ is executed, and then to the
persistent view after executing $\sfence$. Note that until this
occurs, we can guarantee that $\PView{z} = \{0\}$, which trivially
guarantees the persistent invariant.

\vspace{5pt}\noindent\textbf{\emph{{Flush Buffering}}}
\label{sec:flush-buffering}
Our next example is a variation of store buffering (\ref{ex:sb}) and
is used to highlight how writes by different threads on different
locations interact with flushes. Here, thread 1 writes to $x$ and
flushes $y$, while thread 2 writes to $y$ then flushes
$x$.\footnote{Note that the \fl operations here are analogous to the
  \loadC instructions in \ref{ex:sb}.} The writes to $w$ and $z$ are
used to witness whether the flushes in both threads have occurred. The
persistent invariant states that, if both $w$ and $z$ hold $1$ in
persistent memory, then either $x$ or $y$ has the new value (\ie 1) in
persistent memory. 
If both threads perform their $\flushC$ operations, then at least one
must flush value $1$ since a $\fl$ cannot be reordered with a
$\storeC$.

Although simple to state, the proof is non-trivial since it requires
careful analysis of the order in which the stores to $x$ and $y$
occur. In the semantics of Cho \etal~\cite{DBLP:conf/pldi/ChoLRK21},
the $\flushC$ corresponding to the {\em second} $\storeC$
instruction executed synchronises with writes to {\em all}
locations. Thus, for example, if thread $1$'s store to $x$ is executed
after thread $2$'s store to $y$, then the subsequent $\flushC$ in
thread~$1$ is guaranteed to flush the new write to
$y$. 

\begin{figure}[!t]
  \centering
  \noindent
\scalebox{0.95}{
\begin{tabular}{c||c} 
\multicolumn{2}{c}
{ $\assert{\forall o \in \{w, x, y, z\}, \tid \in \{1, 2\}\ldotp \View{o}{\tid} = \PView{o} = \{0\}}$ }\\
$\begin{array}{l}
\assert{
\begin{array}{l}
(\hat{a},\hat{b} = 0, 0 \wedge \PView{z} =\{0\}  ) \vee {} \\
 \left(
  \begin{array}{l}
\hat{a},\hat{b} = 0,1 \wedge \LastR{y}{2} \wedge \\ \View{y}{2} = \{1\}  \wedge \PView{w} =\{0\}
  \end{array}
\right)
\end{array}}\\
\Aux{\store{x}{1}}{\hat{a} := \hat{b} + 1};\\
\assert{
  \begin{array}[t]{l}
    \left(
    \begin{array}{l}
      \hat{a} = 1 \wedge \hat{b} \in \{0,2\} \wedge \\
      (\PView{z} =\{0\} \lor \PView{x} =\{1\}
    \end{array}
    \right)\vee {} \\
    \left(
    \begin{array}{l}
      \hat{a}, \hat{b} = 2, 1 \wedge \LastR{y}{2} \wedge {}\\
      \View{y}{2} = \{1\} \wedge \LastF{y}{1} \wedge \PView{w} = \{0\}
    \end{array}
    \right)
  \end{array}
} \\
   \flush{y};
   \\
   \assert{\begin{array}[t]{l}
             \left(
             \begin{array}{l}
               \hat{a} = 1 \wedge \hat{b} \in \{0,2\} \wedge {} \\
               (\PView{z} =\{0\} \lor \PView{x} =\{1\}))
             \end{array}\right) \vee {} \\
             (\hat{a},\hat{b} = 2,1 \wedge \PView{y} =\{1\}  ) 
           \end{array}}
   \\
   \store{w}{1};
   \\
   \assert{\begin{array}[t]{l}
             \left(
             \begin{array}{l}
               \hat{a} = 1 \wedge \hat{b} \in \{0,2\} \wedge {} \\
               (\PView{z} =\{0\} \lor \PView{x} =\{1\})
             \end{array}\right) \vee {} \\
     (\hat{a},\hat{b} = 2,1 \wedge \PView{y} =\{1\} ) 
   \end{array}}
\end{array}$    
& 
$\begin{array}{l}
\assert{
   \begin{array}[t]{l}
     (\hat{a},\hat{b} = 0,0 \wedge \PView{w} =\{0\})  \vee {} \\
     \left(
     \begin{array}{l}
       \hat{a},\hat{b}=1,0   \wedge \LastR{x}{1} \wedge{} \\
       \View{x}{1} =\{1\} \wedge  \PView{z} =\{0\}
     \end{array}
     \right)
   \end{array}
   }  
   \\
\Aux{\store{y}{1}}{\hat{b} := \hat{a} + 1};\\
\assert{
     \left(
     \begin{array}{l}
       \hat{b}=1 \wedge \hat{a} \in \{0,2\} \wedge {} \\
       (\PView{w} =\{0\} \lor \PView{y} =\{1\})
     \end{array}
   \right) \vee {} \\
   \begin{array}[t]{l}
     \left(
     \begin{array}{l}
       \hat{a},\hat{b}=1,2  \wedge \LastR{x}{1} \wedge \\
       \View{x}{1} =\{1\} \wedge  \LastF{x}{2}
     \wedge  \PView{z} =\{0\}
     \end{array}
\right)
   \end{array}
}
\\ 
 \flush{x} ;
\\
\assert{
   \begin{array}[t]{l}
          \left(
     \begin{array}{l}
       \hat{b}=1 \wedge \hat{a} \in \{0,2\} \wedge {} \\
  (\PView{w} =\{0\} \lor \PView{y} =\{1\}) 
     \end{array}
     \right) \vee {} \\
     (\hat{a},\hat{b}=1,2  \wedge  \PView{x} =\{1\})
   \end{array}
}
\\ 
\store{z}{1}; \\
\assert{
   \begin{array}[t]{l}
               \left(
     \begin{array}{l}
       \hat{b}=1 \wedge \hat{a} \in \{0,2\}   \wedge {} \\
       (\PView{w} =\{0\} \lor \PView{y} =\{1\})
     \end{array}
     \right) \vee {} \\
     (\hat{a},\hat{b}=1,2  \wedge   \PView{x} =\{1\})
   \end{array}
} 
\end{array}$
 \\

\multicolumn{2}{c}
{ $\assert{
\begin{array}{l}
  (\hat{a}, \hat{b} = 1, 2 \wedge \PView{x} =\{1\}) \lor (\hat{a}, \hat{b} = 2, 1 \wedge \PView{y} =\{1\})    
\end{array}}$ }
  \\
\multicolumn{2}{c}
{$\recassert{\PView{w}=\{1\} \wedge \PView{z}=\{1\} \imp \PView{x}=\{1\} \vee \PView{y}=\{1\}}$}
\end{tabular}
}
\vspace{-5pt}
\caption{Proof outline for flush buffering}
\label{fig:po-fb}
\end{figure}

The above intuition requires reasoning about the order in which
operations occur. To facilitate this, we use auxiliary variables
$\hat{a}$ and $\hat{b}$ to record the order in which the writes to $x$
and $y$ occur; $\hat{a}=1$ iff the write to $x$ occurs before the
write to $y$, and $\hat{a}=2$ iff the write to $x$ occurs after the
write to $y$. W.l.o.g., let us now consider the precondition of
$\flush{y}$ (the reasoning for $\flush{x}$ is symmetric). There are
two disjuncts to consider.
\begin{itemize}
\item The first disjunct describes the case in which thread $1$
  executes its store before thread $2$. From here, there is a danger
  that the thread $1$ can terminate having flushed $0$ for
  $y$. However, from this state, thread $2$ is guaranteed to flush~$1$
  for $x$ before setting $z$ to $1$, satisfying the persistent
  invariant, as described by the second disjunct of each assertion in
  thread 2.


\item The second disjunct describes the case in which thread $1$
  executes its store after thread $2$. In this case, thread $1$ is
  guaranteed to flush $1$ for $y$, and this fact is captured by the
  conjunct
  $\LastR{y}{2} \wedge \View{y}{2} = \{1\} \wedge \LastF{y}{1}$, which
  ensures that
  \begin{enumerate*}
  \item thread $2$ sees the last write to $y$; 
  \item the only
    value visible for $y$ to thread $2$ is $1$; and
    
  \item a flush performed by thread $1$ is guaranteed to flush the
    last write to $y$.
  \end{enumerate*}
  Note that by 1) and 2), we are guaranteed that the last write to $y$
  has value $1$. We use these three facts to deduce that
  $\PView{y} = \{1\}$ in the second disjunct of the postcondition of
  $\flush{y}$ using rule ${\sf FP}_3$.
\end{itemize}

\vspace{5pt}\noindent\textbf{\emph{Epoch Persistency}}
\label{sec:epoch-persistency}
In our next example, we demonstrate how writes of different threads on
the same location interact with an optimised flush in the same
location, as well as how the ordering of optimised flushes/loads
alters the persistency behaviour. The crash invariant of
\cref{fig:po-ep2} states that if $z$ and $y$ hold the value $1$ in
persistent memory then $x$ has the value $2$ in persistent memory.

\begin{figure}[!t]
  \centering
\scalebox{0.95}
{
\begin{tabular}{c||c}
\multicolumn{2}{c}
{ $\assert{(\forall \tid \in \{1,2\}, o \in \{x,y,z\}. \View{o}{\tid} =  \PView{o} = \{0\}) \wedge a =0}$ }\\
$\begin{array}{l}
  \assert{
   \begin{array}{l}
     \oc{x}{2} = 0  \wedge {} \\
     \left(
     \begin{array}{l}
       (\View{x}{2} = 0 \wedge \View{x}{1} = 0) \lor \\
       (\View{x}{2} = 1 \wedge \View{x}{1} =\{ 0,1\})
     \end{array}
     \right)
   \end{array}
   } \\
   \store{x}{2};
   \\
   \assert{
   \begin{array}{l}
     \oc{x}{2}=1 \wedge {} \\
     \left(
     \begin{array}{l}
       \left(
       \begin{array}{l}
         \LastR{x}{1} \wedge \View{x}{1} =\{2\} \wedge{}  \\
         \View{x}{2} \subseteq \{1,2 \}
       \end{array}
\right)  
       \lor  \\
       \View{x}{2} \subseteq \{0,1,2 \}
     \end{array}
     \right)
  \end{array}
}
\end{array}$
& 
$\begin{array}{l}
   \assert{\PView{y} = \{0\} \wedge \PView{z} = \{0\} \wedge  (\oc{x}{2} \in \{ 0,1\})}
   \\
   \store{x}{1};
   \\
   \assert{
   \begin{array}{l}
     \left(
     \View{x}{2}=1 \lor \left(\begin{array}{l}
                          \View{x}{2}=\{1,2\} \wedge \oc{x}{2}=1 \wedge {} \\
                          \LastR{x}{1} \wedge \View{x}{1}=2 
     \end{array}\right)
\right) \wedge  \\
     \PView{y} = \{0\} \wedge \PView{z} = \{0\}
   \end{array}
}
   \\
   \load{a}{x};
   \\
   \assert{ (a = 2  \imp \View{x}{2}=\{2\})  \wedge 
   \PView{y} = \{0\} \wedge \PView{z} = \{0\} }
   \\
   \flushopt{x}; \\
   \assert{ (a = 2  \imp  \AView{x}{2} =\{2\})   \wedge 
   \PView{y} = \{0\} \wedge \PView{z} = \{0\}  }
   \\
   \ifC {a = 2} \\
   \qquad \assert{ \AView{x}{2} =\{2\}   \wedge \PView{y} = \{0\} \wedge \PView{z} = \{0\}   }
   \\
   \qquad \store{y}{1};
   \\
   \assert{ (\AView{x}{2} =\{2\}   \lor \PView{y} = \{0\}) \wedge \PView{z} = \{0\}}
   \\
   \sfence;
   \\
   \assert{ \PView{x} = \{2\}   \lor \PView{y} = \{0\}}
   \\
   \store{z}{1};
   \\
   \assert{\PView{x} = \{2\}   \lor \PView{y} = \{0\} \lor \PView{z} = \{0\}}
 \end{array}$
\\
\multicolumn{2}{c}
{ $\assert{
\begin{array}{l}
\PView{x} = \{2\}   \lor \PView{y} = \{0\} \lor \PView{z} = \{0\} \end{array}}$ \qquad \qquad \qquad \qquad \qquad \qquad} 
  \\
  \multicolumn{2}{c}
  {$\recassert{\PView{y}=\{1\} \wedge \PView{z}=\{1\} \Rightarrow \PView{x}=\{2\}}$\qquad \qquad \qquad \qquad \qquad \qquad}
\end{tabular}
}\vspace{-5pt}
  
  \caption{Proof outline for epoch persistency}
  \label{fig:po-ep2}
\end{figure}
In order for thread 2 to read value $2$ for $x$, the $\storeC$ of $2$
at $x$ must be performed before the $\storeC$ of $1$ and
$\View{x}{2}=\{1,2 \}$. Unlike the previous example, establishing the
persistent invariant for thread 2, requires reasoning about the view
of thread 2 for address $x$ (\ie $\View{x}{2}$) after the execution of
the instruction $\load{a}{x}$. Notice here that $\load{a}{x}$ is
ordered with respect to the later $\flushopt{x}$
instruction. Consequently, any impact of the execution of the \ld on
$\View{x}{2}$, will also affect $\AView{x}{2}$. Taking into account
the ordering of the writes at the address $x$, we can conclude that if
thread $2$ reads the value $2$, it reads the value of the last write
at $x$. This is expressed with the assertion $\LastR{x}{1}$ in the
precondition of $\load{a}{x}$, which states that the threads 1's view
of $x$ is the last write to $x$. By rule $\rulename{LP_3}$, if a
thread $\tid$'s view of an address $x$ contains only the last write at
this address, and the last value written at this address appears only
once at the memory, then if a thread $\tid$ read this value at $x$,
its view of $x$ (\ie $\View{x}{\tid}$) is guaranteed to contain only
the last written value at $x$. Consequently, after reading value $2$,
thread 2's view of $x$ contains only the value $2$ (\ie
$\View{x}{2}=\{2\}$). Execution of $\flushopt{x}$ ensures
$\AView{x}{2}$ (by rule $\rulename{OP}$). As a result, in the case
that the {\bf if} statement succeeds, after the execution of the
$\sfence$ it is guaranteed that the value 2 is persisted at $x$ (\ie
$\PView{x} = \{2\}$). In the case that the {\bf if} statement fails,
$\PView{y} = \{0\}$ must hold, thus the persistent invariant holds
trivially. 

\begin{figure}[!t]
{
\begin{mathpar}
\inferrule[{(assign)}]{
\alpha = a := e \\\\
\val = \tstate.\xxrmap(e) \\\\
\tstate' = \tstate[\xxrmap(a) \mapsto \val]
}
{\tup{\tstate,M} \astep{\alpha} \tup{\tstate',M}}
\and
%
\inferrule[{(store)}]{
\alpha = \store{\loc}{e} \\\\
\val = \tstate.\xxrmap(e) \\\\
M' = M \concat [\wmsg{\loc}{\val}] \\\\
\tstate' = \tstate[\xxcoh(\loc) \mapsto \size{M}]
}
{\tup{\tstate,M} \astep{\alpha} \tup{\tstate',M'}}
\and
\inferrule[{(load-internal)}]{
\alpha = \load{a}{\loc} \\\\
M[\tsa] = \wmsg{\loc}{\val} \\\\
\tstate.\xxcoh(\loc) = \tsa \\\\
\tstate' = \tstate[\xxrmap(a) \mapsto \val]
}
{\tup{\tstate,M} \astep{\alpha} \tup{\tstate',M}}
\vspace{-5pt}\\
\inferrule[{(load-external)}]{
\inarrC{
\alpha = \load{a}{\loc} \\
M[\tsa] = \wmsg{\loc}{\val} \\
\tstate.\xxcoh(\loc) < \tsa \\
\noloc{\tstate.\vrp}{\tsa}{M}{\loc}}
\\
\tstate' = \tstate \recupdate{ \begin{array}{@{}l@{}}
\xxrmap(a) \mapsto \val, \\
\xxcoh(\loc) \mapsto \tsa, \\
\vrp \mapstoj \tsa, \\
\vpp \mapstoj \tsa
\end{array} }
}
{\tup{\tstate,M} \astep{\alpha} \tup{\tstate',M}}
\and
\inferrule[{(sfence)}]{\\
\alpha = \sfence \\\\
\tstate' = \tstate \recupdate{ \begin{array}{@{}l@{}}
\vpp \mapstoj \tstate.\xxmaxcoh, \\
\xxper \mapstoj \tstate.\xxpera
\end{array} }
}
{\tup{\tstate,M} \astep{\alpha} \tup{\tstate',M}}
\vspace{-5pt}\\
\inferrule[{(flush)}]{
\alpha = \flush{\loc} \\\\
\tstate' = \tstate \recupdate{ \begin{array}{@{}l@{}}
\xxpera(\loc) \mapstoj \tstate.\xxmaxcoh, \\
\xxper(\loc) \mapstoj \tstate.\xxmaxcoh
\end{array} }
}
{\tup{\tstate,M} \astep{\alpha} \tup{\tstate',M}}
\and
\inferrule[{(flushopt)}]{
\alpha = \flushopt{\loc} \\\\
\tstate' = \tstate[\xxpera(\loc) \mapstoj \tstate.\xxcoh(\loc) \sqcup \tstate.\vpp]
}  
{\tup{\tstate,M} \astep{\alpha} \tup{\tstate',M}}
\vspace{-5pt}\\
%
\noindent\rule{\textwidth}{0.5pt}
\vspace{-5pt}\\
\inferrule[{(program-normal)}]{
\pc(\tid)=i \\
\prog(\tid,i)=\stepgoto{\alpha}{j} \\\\
\tup{\Ts(\tid),M} \astep{\alpha} \tup{\tstate',M'} \\\\
\pc'=\pc[\tid\mapsto j] \\ \Ts' = \Ts[\tid \mapsto \tstate']
}
{ \tup{\pc,\Ts,M,\Gmem} \Rightarrow_\prog \tup{\pc',\Ts',M',\Gmem}}
\hfill
\inferrule[{(program-if)}]{
\pc(\tid)=i \\
\prog(\tid,i)=\ifgoto{B}{j}{k} \\\\
\pc'=\pc \left[  \tid\mapsto {\begin{cases} j & \Ts(\tid).\xxrmap(B)=\true \\ k & 
\Ts(\tid).\xxrmap(B)=\false \end{cases}} \right] 
}
{ \tup{\pc,\Ts,M,\Gmem} \Rightarrow_\prog \tup{\pc',\Ts,M,\Gmem}}
\vspace{-5pt}\\
\inferrule[{(program-ghost)}]{
\pc(\tid)=i \\
\prog(\tid,i)=\Aux{\stepgoto{\alpha}{j}}{\hat{a} := \hat{e}}  \\\\
\tup{\Ts(\tid),M} \astep{\alpha} \tup{\tstate',M'} \\\\
\pc'=\pc[\tid\mapsto j] \\
\Ts' = \Ts[\tid \mapsto \tstate'] \\
\Gmem'=\Gmem[\hat{a}\mapsto \Gmem(\hat{e})]
}
{ \tup{\pc,\Ts,M,\Gmem} \Rightarrow_\prog \tup{\pc',\Ts',M',\Gmem'}}
\end{mathpar}
}\vspace{-10pt}
\caption{Transitions of \viewpintel{}  for a program $\prog$}
\label{fig:full:viewpintel}
\end{figure}

\section{\logic Soundness}
\label{sec:soundness}


In this section we present the \viewpintel{} model from~\cite{DBLP:conf/pldi/ChoLRK21} (\cref{sec:model}),
formally interpret our assertions as predicates on states of that model (\cref{sec:assertions_formal}),
and establish the soundness of the proposed reasoning technique (\cref{sec:soundness_sub}).

\subsection{The \viewpintel{} Model}
\label{sec:model}

Like previous view-based models, \viewpintel{} employs a non-standard
memory capturing all previously executed writes,
alongside with so-called ``thread views'' that track 
several position(s) of each thread in that history and enforce limitations
on the ability of the thread to read from and write to the memory.
In addition, the thread views contain the necessary information
for determining the possible contents of the non-volatile memory upon a system crash.
Formally, \viewpintel{}'s memory and thread states are defined as follows.

\begin{definition}[\viewpintel{}'s memory]{\rm 
A \emph{memory} $M\in \xxMemory$ is a list of \emph{messages},
where each message has the form $\wmsg{\loc}{\val}$
for some $\loc\in\Loc$ and $\val\in\Val$.
We use $\xxmsg.\xxloc$ and $\xxmsg.\xxval$ to refer to the two 
components of a message $\xxmsg$.
We use standard list notations for memories
(\eg $M_1 \concat M_2$ for appending memories,
 $[\xxmsg]$ for a singleton memory,
 and $\size{M}$ for the length of $M$).
We refer to indices (starting from $0$) in a memory $M$ as \emph{timestamps},
and denote the $\tsa$'th element of $M$ as $M[\tsa]$.
We use $\sqcup$ for obtaining the maximum among timestamps
(\ie $\tsa_1 \sqcup \tsa_2 
= \mathrm{max}(\tsa_1,\tsa_2)$),
and extend this notation pointwise to functions.
We write $\noloc{\tsa_1}{\tsa_2}{M}{\loc}$ for the condition
$\forall \tsa_2 < \tsa \leq \tsa_1\ldotp M[\tsa].\xxloc \neq \loc$.
}\end{definition}

\begin{definition}[\viewpintel{}'s thread states]{\rm 
A \emph{thread state} $\tstate \in \xxThread$ is a record consisting of the following fields:
$\xxcoh: \Loc \to \Nats$,
$\vrp : \Nats$,
$\vpp: \mathbb{N}$,
$\xxpera : \Loc \to \Nats$,
and 
$\xxper: \Loc \to \Nats$.
We use standard function/record update notation 
(\eg $\tstate' = \tstate[\xxcoh(\loc) \mapsto \tsa]$ denotes the
            thread state obtained from $\tstate$ be modifying the $\loc$ entry in the $ \xxcoh$ component of $\tstate$ to $\tsa$).
In addition, $\mapstoj$ is used to incorporate certain timestamps in fields
(\eg  $\tstate[\vrp \mapstoj \tsa]$
            denotes the
            thread state obtained from $\tstate$ be modifying  the $ \vrp$ component of $\tstate$ to 
            $\tstate.\vrp \sqcup \tsa$).
We denote by $\tstate.\xxmaxcoh$ the maximum among the coherence view timestamps
($\tstate.\xxmaxcoh = 
\bigsqcup_\loc \tstate.\xxcoh(\loc)$).
}\end{definition}

The two components, together with program counters and the ``ghost memory'', 
are combined in \viewpintel{}'s machine states as defined next.

\begin{definition}[\viewpintel{}'s machine states]{\rm 
A \emph{machine state}  is a tuple $\mstate = \tup{\pc,\Ts,M,\Gmem}$ where
$\pc :\xxTId \to \Label$ is a mapping assigning the next program label to be executed by each thread,
$\Ts :\xxTId \to \xxThread$ is a mapping assigning the current thread state to each thread,
$M\in\xxMemory$ is the current memory,
and $\Gmem :\AuxVar \to \Val$ is storing the current values of the auxiliary variables.
Below we assume that $\Gmem$ is  extended to expressions $\hat{e}\in\AuxExp$ in a standard way.
We denote the components of a machine state $\mstate$ by
$\mstate.\pc$,
$\mstate.\Ts$,
$\mstate.M$, and 
$\mstate.\Gmem$.
In addition, we denote by $\mstate.\xxmaxper(\loc)$ the maximum among the persistency view timestamps
for location $\loc$
($\mstate.\xxmaxper = 
\bigsqcup_\tid \mstate.\Ts(\tid).\xxper(\loc)$).
}\end{definition}

The transitions of \viewpintel{} are presented in
\cref{fig:full:viewpintel}.  These closely follow the model
in~\cite{DBLP:conf/pldi/ChoLRK21} with minor presentational
simplifications.  Note, however, that, for simplicity and
following~\cite{DBLP:journals/pacmpl/KhyzhaL21}, we conservatively
assume that writes persist atomically at the location granularity
(representing, \eg machine words) rather than at the granularity of
the width of a cache line. We refer the interested reader
to~\cite{DBLP:conf/pldi/ChoLRK21} for a detailed discussion of the
transitions rules in \cref{fig:full:viewpintel}. 



The above operational definitions naturally induce a notion of a
execution (or a ``run'') of \viewpintel{} on a certain program $\prog$
starting from some initial state of the form
$\tup {\lambda \tid\ldotp \initlabel, \Ts, M, \Gmem}$.  A system crash
might occur at any point during the execution.  Again, following the
model of~\cite{DBLP:conf/pldi/ChoLRK21}, the non-volatile memory
(\nvm) is not modeled as a concrete part of the state.  Instead, the
possible contents of the \nvm can be inferred from the machine state
(specifically from the memory and the $\xxper$ views of the different
threads), as defined next. This definition is presented as ``crash
transition'' in~\cite{DBLP:conf/pldi/ChoLRK21}.


\begin{definition} \label{def:nvm}
  {\rm 
A non-volatile memory $\nvm:\Loc \to \Val$ is \emph{possible
in a state $\mstate$} if for every $\loc\in \Loc$, 
there exists some $\tsa$ such that 
$\mstate.M[\tsa] = \wmsg{\loc}{\nvm(\loc)}$
and $\noloc{\mstate.\xxmaxper(\loc)}{\tsa}{\mstate.M}{\loc}$.
}\end{definition}


\subsection{The Semantics of \logic Assertions}
\label{sec:assertions_formal}

\newcommand{\OTSF}[1]{{\sf TS^{OF}_{#1}}}
\newcommand{\OTS}[1]{{\sf TS_{#1}}}
\newcommand{\OPTS}{{\sf TS^{P}}}
\newcommand{\OATS}[1]{{\sf TS^{A}_{#1}}}
\newcommand{\CVTS}[1]{{\sf TS^{OV}_{#1}}}
\newcommand{\COBTS}[1]{{\sf TS^{CO}_{#1}}}
\newcommand{\VALS}{{\sf Vals}}
\newcommand{\LAST}{{\sf Last}}

We present the formal definitions of the expressions
introduced in \cref{subsec:assertions} in terms of \viewpintel{}'s machine states.

\smallskip
\noindent\textbf{\emph{Current and conditional views}}
When formalising the {\em current} and {\em conditional view}
expressions, we start with auxiliary functions that return the sets of
observable timestamps visible to the components in question, then
extract the values in memory corresponding these timestamps. To
facilitate this, we define

\smallskip\hfill
$\VALS(M, TS) \eqdef \{M[t].\xxloc \mid t \in TS\}$\hfill{}

\smallskip \noindent
where $M \in \xxMemory$ and $TS$ is a set of timestamps.

\smallskip
\noindent\textbf{\emph{Thread view}}
To define the meaning of the thread view expression,
$\valspred{\loc}{\tid}$, we use:
\begin{align*}
  \OTSF{\tid}(\mstate, \loc, \tsa)  & \defeq \set{ \tsb \st \mstate.M[\tsb].\xxloc = \loc \land \mstate.\Ts(\tid).\xxcoh(\loc) \leq \tsb
                                                     \land \noloc{\tsa}{\tsb}{\mstate.M}{\loc}} \\
  \OTS{\tid}(\mstate, \loc) & \defeq \OTSF{\tid}(\mstate, \loc, \mstate.\Ts(\tid).\vrp) 
\end{align*}

$\OTSF{\tid}(\mstate, \loc, \tsa)$ returns the
set of \emph{timestamps} that are {\em observable from} timestamp
$\tsa$ for thread $\tid$ to read for location $\loc$ in state
$\mstate$; and $\OTS{\tid}(\mstate, \loc)$ returns the set of
\emph{timestamps} that are {\em observable} for $\tid$ to read $\loc$
in $\mstate$. Note that after instantiating $t$ to
$\mstate.\Ts(\tid).\vrp$ in $\OTSF{\tid}(\mstate, \loc, \tsa)$, we
obtain the premises of the load rules in
\cref{fig:full:viewpintel}. Then,
$\valspred{\loc}{\tid} \defeq \lambda \mstate\ldotp \VALS(\mstate.M,
\OTS{\tid}(\mstate, \loc))$,
\ie is the set of values in $\mstate.M$ corresponding to the
timestamps in $\OTS{\tid}(\mstate, \loc)$. 
  
\smallskip
\noindent\textbf{\emph{Persistent memory view}} For the persistent memory view expression, $\PView{\loc}$,
we use: \smallskip

\hfill
$\OPTS(\mstate, \loc) = 
\set{  \tsa \st  \mstate.M[\tsa].\xxloc = \loc  \land 
  \noloc{ \mstate.\xxmaxper(\loc)}{\tsa}{\mstate.M}{\loc}}$\hfill{}

\smallskip\noindent which returns the set of {\em timestamps} that are observable to the
{\em persistent memory} for $\loc$ in $\mstate$. Then, $\PView{\loc}
\defeq \lambda \mstate\ldotp \VALS(\mstate.M, \OPTS(\mstate,
\loc))
$.  Note that the second conjunct within the definition of
$\OPTS(\mstate,
\loc)$ is precisely the condition that links \pxes states to NVM
states (\cref{def:nvm}). Given this definition, we have:
\begin{proposition}
\label{prop:nvm}
A non-volatile memory $\nvm:\Loc \to \Val$ is possible
in a state $\mstate$
iff $\nvm(\loc)\in\PView{\loc}(\mstate)$ 
for every $\loc\in\Loc$.
\end{proposition}

\noindent\textbf{\emph{Asynchronous memory view}}
To define the meaning of the asynchronous memory view,
$\AView{\loc}{\tid}$, we use: \smallskip

\hfill
$    \OATS{\tid}(\mstate, \loc) \defeq \set{\tsa \st \mstate.M[\tsa].\xxloc = \loc \land   
\noloc{ \mstate.\Ts(\tid).\xxpera(\loc)}{\tsa}{\mstate.M}{\loc}}$ \hfill {}

\smallskip \noindent which returns the timestamps of the asynchronous view of thread
  $\tid$ in location $\loc$ and state $\mstate$.  Then, as before, 
  $\AView{\loc}{\tid} \defeq \lambda \mstate\ldotp \VALS(\mstate.M,
  \OATS{\tid}(\mstate, \loc))$. 


\smallskip
\noindent\textbf{\emph{Conditional view}} The 
functions used to define conditional memory view,
$\CView{\loc}{\val}{\loca}{\tid}$, are slightly more sophisticated
than those above. We define:
  \begin{align*}
    \CVTS{\tid}(\mstate,\loc,\val) & \eqdef
      \left\{\begin{array}{@{}l@{~~}|@{~~}l@{}}
               t'  & \exists{\tsa \in \OTS{\tid}(\mstate, \loc)}.\
                     \mstate.M[\tsa].\xxval = \val \land {} \\
                   & \qquad t' =
                     \begin{array}[t]{@{}l@{}}
                       \kwif \tsa = \mstate.\Ts(\tid).\xxcoh(\loc)\ \kwthen\  \mstate.\Ts(\tid).\vrp\\
                       \kwelse\ t \sqcup \mstate.\Ts(\tid).\vrp
                     \end{array}
             \end{array}\right\}
    \\
    \COBTS{\tid}(\mstate,\loc,\val,\loca) & \eqdef    \bigcup \set{\OTSF{\tid}(\mstate, \loca, t) \mid t \in \CVTS{\tid}(\mstate,\loc,\val)}
  \end{align*}
  where $\CVTS{\tid}(\mstate,\loc,\val)$ returns the set of timestamps
  that $\tid$ can observe for $\loc$ with value $\val$. Assuming $\tsa$
  is a timestamp that $\tid$ can observe for $\loc$, and the value for
  $\loc$ at $\tsa$ is $\val$, the corresponding timestamp $t'$ that
  $\CVTS{\tid}(\mstate,\loc,\val)$ returns is $\mstate.\Ts(\tid).\vrp$
  if $\tid$'s coherence view for $\loc$ is $\tsa$, and the maximum of $\tsa$
  and $\mstate.\Ts(\tid).\vrp$, otherwise.  Given this,
  $\COBTS{\tid}(\mstate,\loc,\val,\loca)$ returns the timestamps that
  $\tid$ can observe for $y$, from any timestamp
  $t \in \CVTS{\tid}(\mstate,\loc,\val)$. Finally, the set of
  conditional values is defined by
  $\CView{\loc}{\val}{\loca}{\tid} \defeq \lambda \mstate \ldotp
  \VALS(\mstate.M, \COBTS{\tid}(\mstate,\loc,\val,\loca))$.


\smallskip
\noindent\textbf{\emph{Last view assertions}}
We use the following auxiliary definition:

\smallskip\hfill$\LAST(M,\loc)\defeq\bigsqcup \set{\tsa \st M[\tsa].\xxloc = \loc}$\hfill{}

\smallskip\noindent which returns the timestamp of the last write to
$x$ in $M$.  Then, the last view assertions are given by:
\begin{itemize}
\item $\LastR{\loc}{\tid} \defeq \set{\mstate \st \OTS{\tid}(\mstate,
    \loc) = \set{\LAST(\mstate.M,\loc)} }$, \ie $\tid$'s view of
  $x$ in $\mstate$ {\em is} the last write to $x$ in $\mstate$.
\item $\LastF{\loc}{\tid} \defeq \set{\mstate \st
    \LAST(\mstate.M,\loc) \leq \mstate.\Ts(\tid).\xxmaxcoh \sqcup
    \mstate.\xxmaxper(\loc) }$, \ie the maximum of
  $\tid$'s maximum coherence view and the maximum commit view of
  $\loc$ (over all threads) is beyond the last write to
  $\loc$ in $\mstate$. This means that executing a
  $\flush{x}$ operation in $\tid$ will cause the last write of
  $x$ to be flushed (see {\sc Flush} rule in \cref{fig:full:viewpintel}).
\end{itemize}

\noindent\textbf{\emph{Value count}} Finally, the value count expression is defined as follows: \vspace{2pt}

\hfill$\oc{\loc}{\val} \defeq   \lambda \mstate \ldotp \size{\set{\tsa \st \mstate.M[\tsa]= \wmsg{\loc}{\val}}}$\hfill{}

\subsection{Soundness of \logic}
\label{sec:soundness_sub}

Given the above building blocks, the soundness of the proposed reasoning technique
is stated as follows.

\begin{restatable}[Soundness of \logicname{}]{theorem}{soundness}
  \label{thm:soundness}
  Suppose that a program $\prog$ has a valid proof outline $\tup{\ina, \ann, \inv, \fina}$.
Let $\mstate$ be a state of \viewpintel{} that is reachable in an execution of $\prog$
from some state $\mstate_{\mathsf{init}}$ 
of the form $\tup{\lambda \tid\ldotp \initlabel,\Ts_{\mathsf{init}},M_{\mathsf{init}},\Gmem_{\mathsf{init}}}$
such that  $\mstate_{\mathsf{init}}\in \ina$.
Then, the following hold:
\begin{enumerate}
\item For every $\tid\in\TID$, we have that $\mstate \in \ann(\tid,\mstate.\pc(\tid))$.
\item If $\mstate.\pc(\tid)=\finlabel$ for every $\tid\in\TID$, then $\mstate \in \fina$.
\item Every non-volatile memory $\nvm$ that is possible in $\mstate$
satisfies the crash invariant $\inv$.\end{enumerate} 
\end{restatable}



Finally, it is straightforward to show the soundness of a standard ``auxiliary variable transformation''~\cite{DBLP:journals/acta/OwickiG76}
which removes all auxiliary variables from a program $\prog$ (translating 
each command $\Aux{\stepgoto{\alpha}{j}}{\hat{a} := \hat{e}}$ into
$\stepgoto{\alpha}{j}$) provided that the crash invariant and the final assertion
do not contain occurrences of the auxiliary variables. Indeed, it is easy to
see that the auxiliary memory $\Gmem$ in the operational semantics 
in \cref{fig:full:viewpintel} serves only as an instrumentation,
and does not restrict the possible runs.
(Formally, if $\prog'$ is obtained from  $\prog$ by removing all auxiliary variables
and $\tup{\pc,\Ts,M,\Gmem'}$ is reachable in $\Rightarrow_{\prog'}$ from some initial state,
then $\tup{\pc,\Ts,M,\Gmem}$ is reachable in $\Rightarrow_\prog$ from the same state
for some $\Gmem$.)


\section{Mechanisation}


Perhaps the greatest strength of our development is an integrated
Isabelle/HOL mechanisation providing a fully fledged semi-automated
verification tool for \pxes programs. This mechanisation builds on the
existing work on \owickigries{} for RC11 by Dalvandi et
al~\cite{DBLP:conf/ecoop/DalvandiDDW19,DBLP:journals/darts/DalvandiDDW20}
applying it to the \pxes semantics. We start by encoding the
operational semantics of Cho \etal~\cite{DBLP:conf/pldi/ChoLRK21},
followed by the view-based assertions described in
\cref{subsec:assertions}. Then, we prove correctness of all of the
proof rules for the atomic statements, including those described in
\cref{subsec:proof-rules}. These rules can be challenging to prove
since they require unfolding of the assertions and examination of the
low-level operational semantics and their effect on the views of
different system components.

Once proved, the rules provided are highly reusable, and are key to
making verification feasible. In particular, when showing validity of
a proof outline (\cref{def:outline}), Isabelle/HOL is able to generate
the necessary proof obligations (after some minor interactions), then
{\em automatically} able to find the set of high-level proof rules
needed to discharge each proof obligation via the built-in
sledgehammer tool~\cite{DBLP:conf/cade/BohmeN10}. This facility
enables a high degree of experimentation and debugging of proof
outlines, including the ability to reduce the complexity of assertions
once a proof outline has been validated. 

The base development (semantics, view-based assertions, and soundness
of proof rules) comprise  $\sim$7000 lines of Isabelle/HOL code. With
this base development in place, each example comprises 200--400
lines of code (including the encoding of the program, the annotations,
and the proofs of validity). The entire development took approximately 3
months of full-time work.


\section{Related Work}
The soundness of
\logicname{} is proven relative to the \pxes of Cho
\etal~\cite{DBLP:conf/pldi/ChoLRK21}; there are however other
equivalent models in the
literature~\cite{DBLP:journals/pacmpl/RaadWNV20,DBLP:journals/pacmpl/KhyzhaL21,DBLP:journals/pacmpl/AbdullaABKS21}. While
the original persistent x86 semantics has explicit asynchronous
persist instructions~\cite{DBLP:journals/pacmpl/RaadWNV20}, the
underlying model assumed in this work is the one due to Cho
\etal~\cite{DBLP:conf/pldi/ChoLRK21}, whose persist instructions are
synchronous.  Nevertheless, Khyzha and
Lahav~\cite{DBLP:journals/pacmpl/KhyzhaL21} formally proved that the
two alternatives are equivalent when reasoning about states after
crashes (\eg using our ``crash invariants'').

As mentioned in \cref{sec:intro}, the only existing program logic for persistent programs is POG~\cite{DBLP:journals/pacmpl/RaadLV20}, which (as with \logicname) is a descendent of \owickigries~\cite{DBLP:journals/acta/OwickiG76}. 
\logicname{} goes beyond POG by handling examples that involve $\fo$ instructions, which cannot be
directly verified using POG. 
Raad \etal~\cite{DBLP:journals/pacmpl/RaadLV20} provide a transformation technique to replace certain patterns of $\fo$ and $\sfence$ with $\fl$. 
Specifically, given a program $\prog$ that includes $\fo$ instructions, provided that $\prog$ meets
certain conditions, this transformation mechanism rewrites $\prog$
into an equivalent program $\prog'$ that uses $\fl$ instructions
instead, allowing one to use POG. However, there are three limitations
to this strategy:
\begin{enumerate*}
\item the rewriting is an external mechanism that requires stepping outside the POG logic; 
\item the rewriting is potentially expensive and must be done for every program that includes $\fo$; 
and 
\item the transformation technique is incomplete in that not all programs meet the stipulated conditions (\eg Epoch Persistency 2), and thus cannot be verified using this technique.
\end{enumerate*}
\logicname{} has no such limitations, as we showed in the examples in Section~\ref{sec:example-proofs}. Moreover, POG has no corresponding mechanisation, and developing a mechanisation that also efficiently handles the program transformation for $\fo$ instructions would be non-trivial.

The \owickigries{} method was first applied to non-SC memory
consistency by Lahav \etal~\cite{ogra}.  One way that their approach,
which targets the release/acquire memory model, is different from ours
is that they aim to use standard SC-like assertions; in order to
retain soundness under a weak memory model, they had to strengthen the
standard stability conditions on proof outlines.  
Dalvandi \etal~\cite{DBLP:conf/ecoop/DalvandiDDW19,DBLP:journals/corr/abs-2004-02983}
took a different approach when designing their \owickigries{} logic
for the release/acquire fragment of C11: by employing a more
expressive, view-based assertion language, they were able to stick
with the standard stability requirement.  In our work, we follow
Dalvandi \etal's approach.  However, our assertions are fine-tuned to
cope with the other types of view present in \pxes, such as those
corresponding to the persistent and the asynchronous views.  It is
interesting that some of the principles of view-based reasoning apply
to different memory models, and future work could look at unifying
reasoning across models.

Dalvandi \etal~\cite{DBLP:journals/corr/abs-2004-02983} have developed a deeper integration of their view-based logic using the \owickigries{} encoding of Nipkow and Prensa Nieto~\cite{nipkow_prensa_og} in Isabelle/HOL. Such an integration would be straightforward for \logicname{} too, allowing verification to take place without translating programs into a transition system. This would be much more difficult for POG since \owickigries{} rules themselves are different from the standard encoding in Isabelle/HOL, in addition to the transformation required for $\fo$ instructions discussed above.

The idea of extending Hoare triples with crash conditions first appeared in the work of Chen \etal~\cite{crash_hoare_logic}. However, that work supports neither concurrency nor explicit flushing instructions. 
Related ideas are found in the works of Ntzik \etal~\cite{ntzik+15} and Chajed \etal~\cite{perennial}. 
However, in contrast to \logicname{}, both of these works 
\begin{enumerate*} 
	\item assume sequentially consistent memory, as opposed to a weak memory model such as TSO; 
	\item assume strict persistency (where store and persist orders coincide); and 
	\item assume there is a synchronous $\fl$ operation, which is easier to reason about than the asynchronous $\fo$ operation.
\end{enumerate*}

Besides program logics, there have been other recent
efforts to help programmers reason about persistent programs.  For
instance, Abdulla \etal~\cite{DBLP:journals/pacmpl/AbdullaABKS21} have
proven that state-reachability for persistent x86 is
decidable, thus opening the door to automatic verification of
persistent programs, and Gorjiara
\etal~\cite{DBLP:conf/asplos/GorjiaraXD21} have developed a model
checker for finding bugs in persistent programs.





\bibliographystyle{splncs04}
\bibliography{refs}  

\newpage
\appendix

 \section{Additional examples}

\subsection{Second message passing example with a flush instruction}

 \begin{figure}
\centering
\begin{tabular}{c||c}
\multicolumn{2}{c}
  { $\assert{a = 0 \wedge \forall v \in \{x, y, z\}, \tid \in \{1,2\}.\ \View{v}{\tid} = \PView{v} = \{0\}}$ }\\
$\begin{array}{l}
\assert{\View{y}{2} = \{0\}} \\
\store{x}{42};  \\
\assert{
   \begin{array}[t]{@{}l@{}}
     \View{x}{1} = \{42\} \wedge \\
     \View{y}{2} = \{0\}
   \end{array}
} \\
\store{y}{7};   \\
\assert{\true} \\
\end{array}$
& 
$\begin{array}{l}
\assert{ (7 \in \View{y}{2} \imp \csetvalpred{y}{7}{x}{2} = \{42\} )\wedge 
 \View{y}{2} \subseteq \{0, 7\} 
\wedge \PView{z} = \{0\}} \\ 
\load{a}{y}; \\
\assert{ (a = 7 \imp \View{x}{2} = \{42\})  \wedge \PView{z} = \{0\}} \\
\ifC{a \neq 0} \\ 
\qquad \assert{ \View{x}{2} = \{42\} \wedge [z]_P = \{0\}}\\
\qquad \flush{x};   \\
\qquad \assert{ \PView{x} = \{42\}} \\
\qquad \store{z}{1}; \\
\assert{\PView{z} = \{0\} \lor \PView{x} = \{42\}}
\end{array}$
\\
\multicolumn{2}{c}
{ $\assert{
\begin{array}{l}
\PView{z} = \{0\} \lor \PView{x} = \{42\}\end{array}}$ \qquad \qquad \qquad \qquad \qquad \qquad }
  \\
\multicolumn{2}{c}
{$\recassert{z = \PView{1}  \imp  x = \PView{42}  }$} \qquad \qquad \qquad \qquad \qquad \qquad
\end{tabular}

  \label{fig:ampa}
  \caption{Message passing with a $\fl$ instruction }
\end{figure}

\subsection{Flush buffering with flushopt} 

A flush buffering variation where the $\flush$ instructions are
replaced with $\flushopt$ and $\sfence$ instructions is given in
\cref{fig:aflb}.  Notice here, that because of the reodering that can
occur between $\storeC$ and $\flushopt$ instructions on diffrent
addresses, both the value of $x$ and $y$ can be $0$ in persistent
memory even if the value $1$ is persisted at $w$ and $z$.
\begin{figure}

  \begin{center}
\begin{tabular}{c||c}
\multicolumn{2}{c}
{ $\assert{ \forall v \in \{w, x, y, z\}, \tid \in \{1,2\}.\ \View{v}{\tid} = \{0\} \wedge \PView{v}  = \{0\}}$ }\\
$\begin{array}{l}
   \assert{\PView{z} =\{0\} \wedge  \View{y}{1} \subseteq \{0,1\}}
   \\
   \Aux{x := 1}{\hat{a} := \hat{b} + 1};
   \\
   \assert{\PView{z} =\{0\} \wedge  \View{y}{1} \subseteq \{0,1\}}
   \\
   \flushopt{y};
   \\
   \assert{\PView{z} =\{0\} \wedge  \AView{y}{1} \subseteq \{0,1\}}
   \\
   \sfence;
   \\
   \assert{\PView{z} =\{0\} \wedge   \PView{y} \subseteq \{0,1\}}
   \\
   \store{w}{1}; \\
   \assert{\PView{z} =\{0\} \lor   \PView{y} \subseteq \{0,1\}}  
\end{array}    $
& 
$\begin{array}{l}
   \assert{ \PView{w} =\{0\} \wedge  \View{x}{2} \subseteq \{0,1\} }
   \\
   \Aux{y := 1}{\hat{b} := \hat{a} + 1};
   \\
   \assert{\PView{w} =\{0\} \wedge  \View{x}{2} \subseteq \{0,1\}}
   \\   
   \flushopt{x};
   \\
   \assert{\PView{w} =\{0\} \wedge  \AView{x}{2} \subseteq \{0,1\}}
   \\
   \sfence;
   \\
   \assert{ \PView{w} =\{0\} \wedge   \PView{x} \subseteq \{0,1\}}
   \\
   \store{z}{1};
   \\
   \assert{\PView{w} =\{0\} \lor   \PView{x}  \subseteq \{0,1\} }
\end{array}$
 \\

\multicolumn{2}{c}
{ $\assert{
\begin{array}{l}
 (\PView{z} =\{0\} \lor   \PView{y} \subseteq \{0,1\}) \wedge (\PView{w} =\{0\} \lor   \PView{x} \subseteq \{0,1\} )
\end{array}}$ }
  \\
\multicolumn{2}{c}
{$\recassert{\PView{w}{1} \wedge \PView{z}{1} \imp \PView{x} \in \{0,1\} \vee \PView{y} \in \{0,1\} }$}
\end{tabular}
\end{center}

  \label{fig:aflb}
  \caption{ A flush buffering variation with $\flushopt$ and
 $\sfence$ instructions}  
\end{figure}

\subsection{Epoch persistency}

\begin{figure}[t]
  \centering
  \begin{tabular}{c||c}
\multicolumn{2}{c}
  { $\assert{a=0 \wedge \forall o \in \{x,y,z\}, \tid
  \in \{1,2\}\ldotp \View{o}{\tid} = \PView{o} = \{0\}}$ }\\
$\begin{array}{l}
\assert{\View{x}{2} = \{0\} \lor \View{x}{2} = \{ 1\}}\\
\store{x}{2};\\
\assert{\true}
\end{array}$    
& 
$\begin{array}{l}
   \assert{\PView{y} = \{0\} \wedge \PView{z} = \{0\}}
   \\
   \store{x}{1};
   \\
   \assert{ (\View{x}{2}=\{1\} \lor \View{x}{2}=\{1,2\})   \wedge 
   \PView{y} = \{0\} \wedge \PView{z} = \{0\}}
   \\ 
   \flushopt{x};
   \\
   \assert{
   \begin{array}{l}
     ( \AView{x}{2} = \View{x}{2}=\{1\}  \lor \AView{x}{2}\subseteq\View{x}{2}=\{1,2\})   \wedge {} \\
     \PView{y} = \{0\} \wedge \PView{z} = \{0\}
   \end{array}
   }
   \\ 
   \load{a}{x};
   \\
   \assert{(a = 2  \imp  \AView{x}{2} \subseteq\{1,2\})   \wedge 
   \PView{y} = \{0\} \wedge \PView{z} = \{0\}}
   \\
 \ifC{a = 2}  \\
  \qquad \assert{ \AView{x}{2} \subseteq \{1,2\}   \wedge \PView{y} = \{0\} \wedge \PView{z} = \{0\}} \\
  \qquad     \store{y}{1}; \\
           \assert{ (\AView{x}{2} \subseteq \{1,2\}   \lor \PView{y} = \{0\}) \wedge \PView{z} = \{0\} }\\
 \sfence;\\
 \assert{ \PView{x} \subseteq \{1,2\}    \lor \PView{y} = \{0\} }\\
 \store{z}{1}; \\
 \assert{ \PView{x} \subseteq \{1,2\}    \lor \PView{y} = \{0\} \lor \PView{z} = \{0\}}
\end{array}$
\\
\multicolumn{2}{c}
{ $\assert{
\begin{array}{l}
  \PView{x} \subseteq \{1,2\}    \lor \PView{y} = \{0\} \lor \PView{z} = \{0\} \end{array}}$ \qquad \qquad \qquad \qquad \qquad \qquad }
  \\
  \multicolumn{2}{c}
  {$\recassert{\PView{y}=\{1\} \wedge \PView{z}=\{1\} \Rightarrow \PView{x} \subseteq \{1,2\}}$\qquad \qquad \qquad \qquad \qquad \qquad}

  \end{tabular}
  \vspace{-5pt}

  \caption{Proof outline for second epoch persistency example}
  \label{fig:po-ep1}
\end{figure}

We now consider a second epoch persistency example in
\cref{fig:po-ep1}. The only difference between the example
\cref{fig:po-ep1} and \cref{fig:po-ep2} is the ordering of the
operations \fo and \ld. The write to $z$ in both examples is used to
witness whether the instructions of thread~$2$ has occurred. The crash
invariant of \cref{fig:po-ep1} states that if $z$ and $y$ hold the
value $1$ in persistent memory then $x$ has either the value $1$ or
$2$ in the persistent memory.

The first $\storeC$ of thread $2$, and the \fo that follows, are
performed at the same address, therefore cannot be reordered.  Reading
the value $1$ at $y$ implies that the $\storeC$ of value $2$ at $x$ is
performed before the $\storeC$ of value $1$ at $x$. Otherwise, thread
$2$ would only have the option to read the value $1$ at $x$. Given the
aforementioned store order, the \fo of thread $2$ can view either the
value $1$ or $2$ at $x$. The \ld instruction that follows does not
have any impact on the values that the \fo instruction can view as it
can not be reordered before it. The subsequent \sfenceC ensures that
either $1$ or $2$ is persisted at $x$.
 
In this example, the crash invariant holds due to thread 2. The
initialisation clearly satisfies the precondition of the program.  The
postcondition of the instruction $\store{x}{1}$ obtains two disjuncts
regarding the view of thread 2 at $x$. This disjuction is necessary to
establish the stability of the precondition of threads 2's instruction
$\flushopt{x}$, against threads 1's instruction $\store{x}{2}$. In the
case that $\store{x}{2}$ precedes $\store{x}{1}$, after executing
$\store{x}{1}$, $\View{x}{2}=\{1\}$ holds. This follows by rule
$\rulename{SP_1}$. If the order is flipped, $1 \in \View{x}{2}$ before
the execution of $\store{x}{2}$, and execution of this store results
in a new observable value $2 \in \View{x}{2}$. After the $\fo$ is
executed, $\View{x}{2}$ is translated into an asynchronous view (by
rule $\rulename{OP}$). Specifically, if $\View{x}{2}=\{1\}$ then after
the execution of $\fo$, $\AView{x}{2}=\{1\}$. If
$\View{x}{2}=\{1,2 \}$ then after the execution of $\fo$,
$\AView{x}{2} \subseteq\{1,2\}$ holds. Notice that in order for thread
2 to read value $2$ for $x$, this value should be contained in
$\View{x}{2}$. 
As a result, in the case where the ${\bf if}$ statement succeeds, 
$\View{x}{2}=\{1,2 \}$ and consequently
$\AView{x}{2} \subseteq\{1,2\}$. The execution of $\sfence$ $\sf$
translates the asynchronous view $\AView{x}{2} \subseteq\{1,2\}$ to
the corresponding persistent view $\PView{x} \subseteq\{1,2\}$ (by
rule $\rulename{SFP}$).  In case that the ${\bf if}$ statement fails,
we are certain that $\PView{y} =\{0\}$, thus the crash invariant holds
trivially.


\section{Additional \pxes instructions}

\begin{figure}[!t]
  \centering
\begin{gather*}
\inferrule[{(mfence)}]{
\alpha = \mfence \\\\
\tstate' = \tstate \recupdate{ \begin{array}{@{}l@{}}
\vrp \mapstoj \tstate.\xxmaxcoh, \\
\vpp \mapstoj \tstate.\xxmaxcoh
\end{array} }
}
{\tup{\tstate,M} \astep{\alpha} \tup{\tstate',M}}
\\[10pt]
  \inferrule[{(cas)}]{
\inarrC{
\alpha = \cas{a}{\loc}{e_1}{e_2} \\
\val_1 = \tstate.\xxrmap(e_1) \\
\val_2=\tstate.\xxrmap(e_2) \\
M[\tsa] = \wmsg{\loc}{\val_1} \\
\noloc{\size{M}}{\tsa}{M}{\loc} \\
M' = M \concat [\wmsg{\loc}{ \val_2}]
}\qquad \qquad 
\tstate' = \tstate \recupdate{ \begin{array}{@{}l@{}}
\xxrmap(a) \mapsto \true, \\
\xxcoh(\loc) \mapsto \size{M}, \\
\vrp \mapstoj \size{M}, \\
\vpp \mapstoj \size{M}
\end{array} }
}
{\tup{\tstate,M} \astep{\alpha} \tup{\tstate',M'}}
%
\\[10pt]
\inferrule[{(cas-fail-internal)}]{
  \inarrC{
    \alpha = \cas{a}{\loc}{e_1}{e_2} \\
    M[\tsa] = \wmsg{\loc}{\val} \\
    \tstate.\xxcoh(\loc) = \tsa \\
    (\loc \in M(t..|M|{]} \lor v \neq \tstate.\xxrmap(e_1)) } \qquad
  \tstate' = T\recupdate{ \begin{array}{@{}l@{}}
                            \xxrmap(a) \mapsto \false
\end{array} }
}
{\tup{\tstate,M} \astep{\alpha} \tup{\tstate',M}}
\\[10pt]
\inferrule[{(cas-fail-external)}]{
  \inarrC{
    \alpha = \cas{a}{\loc}{e_1}{e_2} \\
    M[\tsa] = \wmsg{\loc}{\val} \\
    \tstate.\xxcoh(\loc) < \tsa \\
    (\loc \in M(t..|M|{]} \lor v \neq \tstate.\xxrmap(e_1)) } \qquad 
  \tstate' = T\recupdate{ \begin{array}{@{}l@{}}
                            \xxrmap(a) \mapsto \false, \\
                              \xxcoh(\loc) \mapsto t, \\
                            \vrp \mapstoj t, \\
                            \vpp \mapstoj  t
\end{array} }
}
{\tup{\tstate,M} \astep{\alpha} \tup{\tstate',M}}
\end{gather*}
  
  \caption{Additional instructions}
  \label{fig:additional-instr}
\end{figure}



\subsection{Example with $\casC$}

\begin{figure}[!t]
  \centering
{\small 
\begin{tabular}{c||c}
	\multicolumn{2}{c}
	{ $\assert{ \forall v \in \{lx, x, y, z\}, \tid \in \{1,2,3\}.\ \View{v}{\tid} = \{0\} \wedge \PView{v}  = \{0\}}$ }\\
	$\begin{array}{l}
	\assert{ (\LastRV{lx}{0} \wedge a_1 = 0)  \lor {} \\(\LastRV{lx}{2} \wedge a_1 = 0) }
	\\
	\cas{a_1}{lx}{0}{1} ;
	\\
	\assert{ (a_1 = 1 \wedge \LastRV{lx}{1}) \lor a_1 = 0}
	\\
	\ifC{a_1 = 1}  
	\\
	\assert{\LastRV{lx}{1}}
	\\
	\qquad     \store{x}{1}; \\
	\qquad       \assert{
           \begin{array}[t]{@{}l@{}}
             \LastRV{lx}{1} \wedge{} \\
             \View{x}{1} =\{1\} \wedge \LastR{x}{1}
           \end{array}
           }\\
           \qquad     \store{y}{1}; \\
	\qquad    \assert{
           \begin{array}[t]{@{}l@{}}
             \LastRV{lx}{1} \wedge \View{x}{1} =\{1\} \wedge {}
             \\
             \LastR{x}{1} \wedge \View{y}{1} =\{1\}
           \end{array}
}\\ 
	\qquad     \flush{x}; \\  
           \qquad    \assert{ \LastRV{lx}{1} \wedge \PView{x}=\{1\} \wedge \\
           \LastR{x}{1} \wedge \View{y}{1} =\{1\} \\ \wedge \View{x}{1}=\{1\}  }\\   
	\qquad     \flush{y} ;\\  
	\qquad    \assert{ \LastRV{lx}{1} \wedge \PView{x}=\{1\} \wedge {} \\ \LastR{x}{1} \wedge \PView{y}=\{1\} \wedge\\ \View{x}{1}=\{1\}  }\\     
	
	\qquad \store{lx}{0}; \\
	\assert{ \left(
           \begin{array}{@{}l@{}}
            \PView{x}=\{1\}  \wedge \PView{y}=\{1\}
           \end{array}
           \right) \\ \lor a_1 = 0 }\\

\end{array}    $
& 
$\begin{array}{l}
	\assert{ \left(
   \begin{array}{@{}l@{}}
     \LastRV{lx}{0} \wedge  1 \notin  \View{x}{2} \wedge {} \\  \PView{z} = \{0\} \wedge a_2 = a_3=0
   \end{array}
\right) \lor \\
		
		\left(
   \begin{array}{@{}l@{}}
     \LastRV{lx}{0} \wedge {} \View{x}{1} = \PView{x}= \PView{y}=\{1\} \wedge {}  \\  \LastR{x}{1} \wedge \PView{z} = \{0\}  \wedge a_2=  a_3=0 
   \end{array}
   \right)   \lor \\
		
		(\LastRV{lx}{1} \wedge \PView{z} = \{0\} \wedge a_2 = a_3 = 0 )}
	\\
	\cas{a_2}{lx}{0}{2} ;
	\\
	\assert{ (\LastRV{lx}{2} \wedge a_2 = 1 \wedge  1 \notin \View{x}{2} \wedge \PView{z} = \{0\} ) \lor \\
		
		\left(
   \begin{array}{@{}l@{}}
     \LastRV{lx}{2} \wedge  \PView{x}=   \PView{y}=\{1\} \wedge {} \\ 
     \PView{z} = \{0\}  \wedge a_2=1
     \wedge \View{x}{2}=\{1\}
   \end{array}
\right)   \lor \\
		
		( a_2 = 0 \wedge \PView{z} = \{0\})}
	
	\\   
	\ifC{a_2 = 1}  
	\\
	\qquad \assert{ (\LastRV{lx}{2} \wedge  1 \notin \View{x}{2} \wedge \PView{z} = \{0\} ) \lor \\
		
		(\LastRV{lx}{2} \wedge  \PView{x}=\{1\}\wedge  \PView{y}=\{1\} \wedge  \\ \PView{z} = \{0\} 
		\wedge \View{x}{2}=\{1\} )  }
	\\
	\qquad \load{a_3}{y};
	\\
	\qquad \assert{\left(
   \begin{array}{@{}l@{}}
     a_3 = 1 \imp  \LastRV{lx}{2} \wedge  \PView{x}=\{1\}\wedge {} \\  \PView{y}=\{1\} 
     \wedge \View{x}{2}=\{1\}
   \end{array}
\right)  \wedge {} \\  \LastRV{lx}{2} \wedge  \PView{z} = \{0\}}  \\
	
	\qquad  \ifC{a_3 = 1}  
	\\
	\qquad \qquad \assert{ \PView{x} =\{1\} \wedge   \PView{y} =\{1\}  \wedge \LastRV{lx}{2} }
	\\
	\qquad \qquad \  \store{z}{1}; \\ 
	\qquad \qquad\assert{ \PView{x} =\{1\} \wedge   \PView{y} =\{1\}  \wedge {} \\ \LastRV{lx}{2}  \wedge  \View{z}{1} =\{1\}  }\\
	
	\qquad \qquad \ \flush{z}; \\  
	
	\qquad \assert{( \PView{x} =\{1\} \wedge   \PView{y} =\{1\}  \wedge \LastRV{lx}{2} \\ \wedge  \PView{z} =\{1\}) \\
		\lor ( \PView{z} =\{0\} \wedge  \LastRV{lx}{2}   )}
	
	\\
	\qquad \store{lx}{0}; \\
	\assert{( \PView{x} =\{1\} \wedge   \PView{y} =\{1\}   \wedge  \PView{z} =\{1\}) \\
		\lor ( \PView{z} =\{0\}   )}
\end{array}$
\\

\multicolumn{2}{c}
{ $\assert{
		\begin{array}{l}
		\PView{z} =\{0\} \vee  ( \PView{x} =\{1\} \wedge \PView{y} = \{1\} ) 
		\end{array}}$ }
\\
\multicolumn{2}{c}
{$\recassert{\PView{z} = \{1\}  \imp (\PView{x} = \{1\} \wedge \PView{y} = \{1\} ) }$}
\end{tabular}
}
  
  \caption{CAS-based locking}
  \label{fig:CAS}
\end{figure}

Our next example use the compare and set ($\casC$) instruction. Notice that from the semantics of $\casC$ given in \cref{fig:additional-instr} it can be inferred that a $\casC$ on  address $x$ ( $\cas{a}{\loc}{e_1}{e_2} $) succeeds only if the value that is read in $x$ is $e_1$, and $e_1$ is the last written value on $x$. In order to facilitate reasoning about $\casC$  we introduce another view expression, namely   $\LastRV{x}{v}$. This expression is boolean-valued and holds  if the last written value on $x$ is $v$.  

\cref{fig:proof-rules-pre-post-cas}  and \cref{fig:proof-rules-stable-cas} present a selection of rules for atomic statements and view-based assertions regarding $\casC$. The selection is based on the rules that are used in the proof outline of the example \cref{fig:CAS}. Each of these rules have been proved sound with respect to view-based semantics in Isabelle. For these rules the same conventions are made as for the rules in  \cref{fig:proof-rules-pre-post}  and \cref{fig:proof-rules-stable}. Rule $\rulename{CP_1}$ states that after 
the execution of $\casC$ on $x$ either $a =1$ (indicating that $\casC$  succeeded) and the last written value on $x$ is $e_2$, or $a =0$ (indicating that  $\casC$ failed). By  $\rulename{CP_2}$, providing that $ x \neq y$ and 
$\tid$'s view of $y$ is the set of values $S$ then in the postcondition $\tid$'s view of $y$ is a subset
of $S$. By rule  $\rulename{CP_3}$, given that the last written value on $x$ is $v$ then in the postcondition the last written value either remains the same (indicating that the $\casC$ failed) or $\LastRV{x}{e_2}$.
Rule $\rulename{CP_4}$ states that given $x \neq y$ if  $\tid'$'s view of y is  the last write on $y$ and  the value of this write is  $v$ then in the postcondition either the  $\casC$ succeeds, so $a=1$ and $\tid$'s view of y is $\View{y}{\tid} = \{v\}$ or the $\casC$ fails and $a=0$. By rule $\rulename{CP_5}$ ifthe last written value on $x$ is different from
$e_1$ then $a=0$. By rule $\rulename{SP_8}$ after executing a store on $x$ ($\store{x}{v}$), the value of the last write on $x$ is updated to $v$. We also prove the stability
of several assertions regarding $\casC$ (see \cref{fig:proof-rules-stable-cas} for a selection).

\begin{figure}[t]
  \centering
  \begin{array}[t]{r@{~}c@{~}l@{~}|@{~}l@{~}|@{~}l}
    \textrm{Precondition} & \textrm{Statement} & \textrm{Postcondition} & \textrm{Const.} & \textrm{Ref.}  \\
    \hline  
    \assert{true} & \multirow{5}{*}{$\cas{a}{\loc}{e_1}{e_2} $} & \assert{(a = 1 \wedge \LastRV{x}{e_2}) \lor a = 0  } & & \rulename{CP_1}
    \\
    \assert{\View{y}{\tid'} = S} &  & \assert{\View{y}{\tid'} \subseteq S}  & x \neq y & \rulename{CP_2}
    \\
      \assert{\LastRV{x}{v} } &  & \assert{\LastRV{x}{e_2} \lor \LastRV{x}{v} }  & & \rulename{CP_3}
     
         \\
      \assert{\LastR{y}{\tid'} \wedge \View{y}{\tid'}= \{v\}} &  & \assert{ (a=1 \wedge \View{y}{\tid}= \{v\}) \lor a=0 }  & x\neq y& \rulename{CP_4}
      \\
      \assert{\LastRV{x}{v} } &  & \assert{a=0}  & v \neq e_1 & \rulename{CP_5}
    \\
    \hline    
    \assert{true} & \multirow{1}{*}{$\store{x}{v}$ }  & \assert{\LastRV{x}{v} } 
       & & \rulename{SP_8}
  \end{array}
  \caption{Selected proof rules for atomic statements executed by
    thread $\tid$ regarding $\casC$. Note $\tid$ may be equal to $\tid'$ unless
    explicitly ruled out.}
\label{fig:proof-rules-pre-post-cas}
\end{figure}

\begin{figure}[t]
  \centering
  
  \scalebox{0.97}{
  \begin{array}[t]{@{}l@{~}||@{~}l@{}}
    \begin{array}[t]{l@{~}|@{\,}l@{\,}|@{\,}l@{\,}|@{\,}l}
    \textrm{Statement} & \textrm{Stable Assert.}  & \textrm{Const.} & \textrm{Ref.}  \\
      \hline
      \multirow{1}{*}{$\load{a}{x}$ } 
    & \assert{\LastRV{y}{v}} & & \rulename{LS_6} 

      \\
      \hline
            \multirow{1}{*}{$\flush{x}$ } 
    & \assert{\LastRV{y}{v}}  & & \rulename{FS_6} 
    \\
      \hline
                \multirow{1}{*}{$\store{x}{v}$ } 
 & \assert{\LastRV{y}{v}}  &  x \neq y   & \rulename{WS_8} 
  
    \end{array}
                                                                    &
    \begin{array}[t]{l@{\,}|@{\,}l@{\,}|@{\,}l@{\,}|@{\,}l}
    \textrm{Statement} & \textrm{Stable Assert.}  & \textrm{Const.} & \textrm{Ref.}  \\
    \hline
                    \multirow{4}{*}{$\cas{a}{\loc}{e_1}{e_2} $ } 
      &
          \assert{ v \notin \View{y}{\tid'}  }  & x \neq y& \rulename{CS_1} 
    \\
    & \assert{\LastRV{x}{v} }  & v \neq e_1 & \rulename{CS_2} 
    \\
    & \assert{\PView{y}  = S}  & x \neq y& \rulename{CS_3} 
      \\
    & \assert{\LastR{y}{\tid'} }  & x \neq y & \rulename{CS_4}

    \end{array}
  \end{array}}
  \caption{Selection of stable assertions for atomic statements
    executed by thread $\tid$ regarding $\casC$. Note $x$ may be equal to $y$ and $\tid$
    may be equal to $\tid'$ unless explicitly ruled out.  }
  \label{fig:proof-rules-stable-cas}
\end{figure}

Let us consider now the program of \cref{fig:po-ep2}. In this example we use  $\casC$ as a lock, in order to control accesses on $x$. The  crash invariant here, states that if $z$ holds the value 1 in persistent memory then
$x$ and $y$ should also obtain the value $1$ in persistent memory. In this example the invariant is establised by thread 2. 

In order for the invariant to hold, we must ensure that the $\flushC$
instructions of thread $1$, are executed before thread 2 executes
$\store{z}{1}$. The $\casC$ instructions in the beginning of the two
threads program ensure that the threads are not executing in
parallel. In particular, in order for thread $1$'s $\casC$ to succeed
the value of the last write on $x$ should be $0$. If the value of the
last write on $x$ is 2, it means that thread $2$'s $\casC$ is
executed and the execution point hasn't reached yet the thread $2$'s
instruction $\store{lx}{0}$. In this case the thread 1 $\casC$
fails, and its execution stalls. More concretely, by rule
$\rulename{CP_1}$, after the execution of $\casC$ in thread 1, we can
obtain that either $a_1 = 1 \wedge \LastRV{lx}{1}$ (indicating that
the $\casC$ succeed) or $a_1 = 0$. Respectively, in order for thread
$2$'s $\casC$ to succeed the value of the last write on $x$ should be
$0$. If the value of the last write on $x$ is 1, it means that that
thread $1$'s $\casC$ is executed and the execution point hasn't reached
yet the thread $1$'s instruction $\store{lx}{0}$. In this case the
execution of thread 2 stalls. There are two ways for $\LastRV{lx}{0}$
to hold for thread 2 before the execution of $\casC$. Either the
$\casC$ reads the initial value of $lx$ or it reads the value that
$lx$ obtains after thread 1 executes the instruction
$\store{lx}{0}$. In the second case, which is the desirable one, we
are sure that before thread 2 executes $\casC$,
$ \PView{x} = \PView{y} = \View{y}{2} = 1$. Those cases are described
in the precondition of $\casC$ in thread 2.

The first disjunct of the precondition concerns the case in which
thread $2$'s write on $lx$ is not executed yet, but the value of the
last write on $lx$ is 0. From this it can be inferred that $lx$
obtains is initial value. In this case we are sure that
$1 \notin \View{x}{2}$. The consecutive $\casC$ might succeed,
although it is certain that thread 2 can not read $1$ at $x$. As a
result, the second {\bf if} statement of thread 2 fails, thus
$\PView{z} \neq \{1\}$, consequently the invariant holds.

The second disjunct of the precondition concerns the case in which
thread~$2$'s $\store{lx}{0}$ has been executed.  Because the store of
1 at $x$ by thread 1 is ordered before the store of 0 at $lx$, it is
certain that at this point of execution
$\LastR{x}{1} \wedge \View{x}{1} = \{1\}$ holds. By rule
$\rulename{CP_4}$ if the consecutive $\casC$ succeeds, thread $1$'s
view of x is transferred to thread 2. As a result the {\bf if}
statement that follows succeeds and $\flush{z}$ persists the value 1
at $z$. Because either $\PView{x} = \{1\} \wedge \PView{y} = \{1\}$ or
$\PView{z} = \{1\} $ for every state of thread $2$'s program, the
invariant holds.

The third disjunct of the precondition concerns the case in which
thread $1$'s $\casC$ has succeeded and thus $\LastRV{x}{1}$. In this
case thread $2$'s $\casC$ can not succeed and the invariant holds
trivially.

\begin{figure}[!t]
  \centering
  \noindent
\scalebox{0.95}{
\begin{tabular}{c||c} 
\multicolumn{2}{c}
{ $\assert{\forall o \in \{ x, y\}, \tid \in \{1, 2\}\ldotp \View{o}{\tid}  = \{0\}}$ }\\
$\begin{array}{l}
\assert{
\begin{array}{l}
(\hat{a},\hat{b} = 0, 0 \wedge  \hat{d} = 0  ) \vee {} \\
 \left(
  \begin{array}{l}
\hat{a},\hat{b} = 0,1 \wedge \LastR{y}{2} \wedge \\ \View{y}{2} = \{1\} 
  \end{array}
\right)
\end{array}}\\
\Aux{\store{x}{1}}{\hat{a} := \hat{b} + 1};\\
\assert{
  \begin{array}[t]{l}
    \left(
    \begin{array}{l}
      \hat{a} = 1 \wedge \hat{b} \in \{0,2\} \wedge \\
      (\hat{d} = 0 \lor  r_2 = 1
    \end{array}
    \right)\vee {} \\
    \left(
    \begin{array}{l}
      \hat{a}, \hat{b} = 2, 1 \wedge \LastR{y}{2} \wedge {}\\
      \View{y}{2} = \{1\} \wedge \LastM{y}{1} \wedge \hat{d} = 0
    \end{array}
    \right)
  \end{array}
} \\
   \Aux{\mfence}{\hat{c} :=  1};
   \\
   \assert{\begin{array}[t]{l}
             \left(
             \begin{array}{l}
               \hat{a} = 1 \wedge \hat{b} \in \{0,2\} \wedge {} \\
               (\hat{d} = 0 \lor r_2 = 1  ))
             \end{array}\right) \vee {} \\
             (\hat{a},\hat{b} = 2,1 \wedge \View{y}{1} =\{1\}  ) 
           \end{array}}
   \\
   \load{r_1}{y};
   \\
   \assert{\begin{array}[t]{l}
             \left(
             \begin{array}{l}
               \hat{a} = 1 \wedge \hat{b} \in \{0,2\} \wedge {} \\
               (\hat{d} = 0 \lor r_2 = 1)
             \end{array}\right) \vee {} \\
     (\hat{a},\hat{b} = 2,1 \wedge r1 = 1 ) 
   \end{array}}
\end{array}$    
& 
$\begin{array}{l}
\assert{
   \begin{array}[t]{l}
     (\hat{a},\hat{b} = 0,0 \wedge  \hat{c} = 0 )  \vee {} \\
     \left(
     \begin{array}{l}
       \hat{a},\hat{b}=1,0   \wedge \LastR{x}{1} \wedge{} \\
       \View{x}{1} =\{1\} 
     \end{array}
     \right)
   \end{array}
   }  
   \\
\Aux{\store{y}{1}}{\hat{b} := \hat{a} + 1};\\
\assert{
     \left(
     \begin{array}{l}
       \hat{b}=1 \wedge \hat{a} \in \{0,2\} \wedge {} \\
       (\hat{c} = 0 \lor r_1 = 1 )
     \end{array}
   \right) \vee {} \\
   \begin{array}[t]{l}
     \left(
     \begin{array}{l}
       \hat{a},\hat{b}=1,2  \wedge \LastR{x}{1} \wedge \\
       \View{x}{1} =\{1\} \wedge  \LastM{x}{2}
     \wedge  \hat{c}=0  
     \end{array}
\right)
   \end{array}
}
\\ 
  \Aux{\mfence}{\hat{d} :=  1};
\\
\assert{
   \begin{array}[t]{l}
          \left(
     \begin{array}{l}
       \hat{b}=1 \wedge \hat{a} \in \{0,2\} \wedge {} \\
  (\hat{c} = 0  \lor \PView{y} =\{1\}) 
     \end{array}
     \right) \vee {} \\
     (\hat{a},\hat{b}=1,2  \wedge  \View{x}{2} =\{1\})
   \end{array}
}
\\ 
   \load{r_2}{x}; \\
\assert{
   \begin{array}[t]{l}
               \left(
     \begin{array}{l}
       \hat{b}=1 \wedge \hat{a} \in \{0,2\}   \wedge {} \\
       (\hat{c} = 0  \lor r_1 = 1 )
     \end{array}
     \right) \vee {} \\
     (\hat{a},\hat{b}=1,2  \wedge    r_2 = 1)
   \end{array}
} 
\end{array}$
 \\

\multicolumn{2}{c}
{ $\assert{
\begin{array}{l}
  (r_1 = 1 \lor r_2 = 1)    
\end{array}}$ }
  \\
\multicolumn{2}{c}

\end{tabular}
}
\caption{Proof outline for flush buffering}
\label{fig:po-sb}
\end{figure}

\subsection{Example with $\mfenceC$}

Our next example use the $\mfenceC$ instruction. In order to facilitate reasoning about $\mfence$  we introduce another view expression,  $\LastM{x}{\tid}$. This expression is boolean-valued and holds  iff after performing an \mfence operation, the view of thread $\tid$ will be the last write on $x$. Specifically,
\[\LastM{\loc}{\tid} \defeq   \set{\mstate \st 
LAST(\mstate.M,\loc) \leq 
\mstate.\Ts(\tid).\xxmaxcoh
}.
\]
 \cref{fig:proof-rules-pre-post-mfence} and \cref{fig:proof-rules-stable-mfence} extends the proof rules with rules regarding  $\mfenceC$.
 The proof outline of  \cref{fig:po-sb}  follows closely the proof outline of the flush buffering example (\cref{fig:po-fb}). Instead of the assertion  $\LastF{x}{\tid}$  we use the analogous assertion for $\mfenceC$, 
  $\LastM{x}{\tid}$. Two additional auxiliary variables are used to indicate if the $\mfenceC$, of thread 1 (resp. thread 2) is executed. In the end of the execution either $a_1=1$ or $a_2=1$.


\begin{figure}[!t]
  \centering
  
  \begin{array}[t]{r@{~}c@{~}l@{~}|@{~}l@{~}}
    \textrm{Precondition} & \textrm{Statement} & \textrm{Postcondition}  & \textrm{Ref.}  \\
    \hline    
     \assert{true} &  \multirow{1}{*}{$\store{x}{v}$ }  & \assert{\LastM{x}{\tid} }  & \rulename{SP_9} 
 
    \\
    
\hline
\assert{\begin{array}[c]{@{}l@{}}
 \LastR{x}{\tid'} \wedge \LastM{x}{\tid} \wedge {} \View{x}{\tid'} = \{u\}  
    
 \end{array}}
                 &
                  \multirow{2}{*}{\mfence}
& \assert{\valspred{x}{\tid} = \{u\}}  &  \rulename{MFP_1} 
    \\
    \assert{\valspred{x}{\tid}  = S} & 
                   & \assert{\valspred{x}{\tid} \subseteq S} & \rulename{MFP_2}

  \end{array}
  \caption{Selected proof rules for atomic statements executed by
    thread $\tid$ regarding $\mfence$. Note $\tid$ may be equal to $\tid'$ unless
    explicitly ruled out.}
\label{fig:proof-rules-pre-post-mfence}
\end{figure}

\begin{figure}[!t]
  \centering
  
  \fontsize{8.7}{10}\selectfont
    \begin{array}[t]{l@{~}|@{\,}l@{\,}|@{\,}l@{\,}|@{\,}l}
    \textrm{Statement} & \textrm{Stable Assert.}  & \textrm{Const.} & \textrm{Ref.}  \\
      \hline
      \multirow{1}{*}{$\load{a}{x}$ } 
      
                       & \assert{\LastM{y}{\tid'}}   &  & \rulename{LS_7} 

      \\
      \hline
      \multirow{1}{*}{$\store{x}{v}$ }  & \assert{\LastM{y}{\tid'}} & & \rulename{WS_9}
      \\      
   \hline \mfence
    & \assert{\PView{x}  = S} && \rulename{MFS_1} 
    \\
    & \assert{\LastR{x}{\tid'}} && \rulename{MFS_2} 
     \\
     & \assert{\LastM{x}{\tid'}}  & \tid \neq \tid'& \rulename{MFS_3} 
    \\
    &  \assert{\oc{x}{v}= n} && \rulename{MFS_4} 
    \end{array}
   
  \caption{Selection of stable assertions for atomic statements
    executed by thread $\tid$ regarding $\mfence$. Note $x$ may be equal to $y$ and $\tid$
    may be equal to $\tid'$ unless explicitly ruled out.  }
  \label{fig:proof-rules-stable-mfence}
\end{figure}



\section{Proof of Theorem~\ref{thm:soundness}}

\soundness*

\noindent Formally, when checking if $\nvm$ satisfies $\inv$, one
has to translate every $\PView{\loc}$ expression in $\inv$ to
$\{\loc\}$. For example, $\inv=\PView y = \{1\} \imp \PView x = \{1\}$
is translated into $\{y\} = \{1\} \imp \{x\} = \{1\}$, and
$\nvm: \Loc \to \Val$ satisfies $\inv$ if $\nvm(y)=1\imp \nvm(x)=1$
holds. Recall that we assume that the only ``specialised'' logical
expressions in persistent invariants are of the form $\PView{\loc}$.
\begin{proof}
We prove the item first by induction on the length of the trace.
The basis of the induction follows from the {\sf Initialisation} condition in the definition
of a valid outline (\cref{def:outline}).
Now, for each step in the trace performed by thread $\tida$
(\ie a transition obtained by \textsc{program-normal} or \textsc{program-if} with $\tid:=\tida$),
the inductive step follows from the  {\sf Local correctness} condition for $\tid=\tida$,
or for the {\sf Stability} condition for $\tid\neq\tida$.
The second item follows form the first using the {\sf Finalisation} condition.
Finally, to see that the third item holds, let $\nvm:\Loc\to\Val$ be a non-volatile memory
that is possible in $\mstate$.
By the {\sf Persistence} condition, we know that there exists $\tid \in \TID$ 
such that $\ann(\tid, \mstate.\pc(\tid)) \imp \inv$.
By the first item, we have $\mstate \in \ann(\tid,\mstate.\pc(\tid))$,
and so, it follows that $\mstate \in \inv$.
The fact that $\nvm$ satisfies $\inv$ 
satisfies the persistent invariant $\inv$ follows from \cref{prop:nvm}.
\qed
\end{proof}


\end{document}